\documentclass[12pt]{article}
\usepackage{amssymb,amsmath,epsfig}

\begin{document}

\title{\bf Stability of Thin-Shell Wormholes from Regular ABG Black Hole}
\author{M. Sharif \thanks {msharif.math@pu.edu.pk} and Saadia Mumtaz
\thanks{sadiamumtaz17@gmail.com}\\
Department of Mathematics, University of the Punjab,\\
Quaid-e-Azam Campus, Lahore-54590, Pakistan.}

\date{}
\maketitle

\begin{abstract}
In this paper, we construct thin-shell wormholes from regular
Ayon-Beato and Garcia black hole by employing cut and paste
formalism and examine their stability. We analyze attractive and
repulsive characteristics of the respective thin-shell wormholes. A
general equation of state is assumed as a linear perturbation to
explore stability of these constructed wormholes with and without
cosmological constant. We consider linear, logarithmic and Chaplygin
gas models for exotic matter and evaluate stability regions for
different values of charge. It is found that generalized Chaplygin
gas model provides maximum stable regions in de Sitter background
while modified generalized Chaplygin gas and logarithmic gas yield
maximum stable regions in anti-de Sitter spacetime.
\end{abstract}

\section{Introduction}

The stability of wormholes under perturbations has been a
challenging issue for physicists in general relativity. A
``wormhole" is like a hole with two ends associating distant regions
of the universe which takes much less time for traveling as compared
to normal travel time. Besides the lack of observational evidence,
wormholes are considered to be interconvertible with black holes
(BH) \cite{2}. The Einstein-Rosen bridge is an example of
non-traversable wormhole in which wormhole throat shrinks leading to
the existence of event horizon \cite{2a}. Traversable wormholes have
no event horizon which allow the two way observer's motion without
any hindrance \cite{3}.

The existence of exotic matter at the wormhole throat gives rise to
the violation of null (NEC) and weak energy conditions (WEC) which
is an important ingredient for traversable wormholes. The weakest
one is NEC whose violation leads to the violation of WEC and strong
energy conditions. Israel thin-shell formalism is the most useful
technique to compute presuure and energy density which idetifies
exotic matter at thin-shell \cite{4}. Thin-shell wormholes belong to
one of the wormhole classes in which this matter is restricted at
the shell. The unavoidable amount of exotic matter is a debatable
issue for physical viability of thin-shell wormholes. This amount
can be quantified by the volume integral theorem for shell which is
consistent with the concept that a small quantity of exotic matter
is required to support wormhole \cite{5}. Visser \cite{6} proposed
cut and paste procedure for polyhedral wormholes which could make
infinitesimally small contribution of exotic matter such that a
traveler encountering the wormhole does not feel any tidal force.

Thin-shell wormholes can be constructed from the family of regular
BHs which are static spherically symmetric and asymptotically flat
regions with regular (singularity-free) centers. The horizons are
indeed formed, while singularities are avoided in the interior of
BHs. We can choose regular BHs for wormhole construction due to
their viability in high energy collisions. Bardeen was the first,
who proposed a regular BH solution with specific mass to charge
ratio \cite{6a}. Ayon-Beato and Garcia \cite{6b} found another
regular BH coupled with nonlinear electrodynamics known as
Ayon-Beato and Garcia (ABG) BH. Hayward \cite{6c} proposed a similar
type of regular BH which can be reduced to de Sitter and
Schwarzschild spacetimes in the limit $r\rightarrow0$ and
$r\rightarrow\infty$, respectively. Recently, Halilsoy \emph{et al.}
\cite{6d} studied stability of regular Hayward wormhole
configurations.

It is well-known that the implementation of linear perturbations or
equation of state (EoS) has remarkable significance in the
investigation of wormhole stability. In this context, Kim and Lee
\cite{7} analyzed the role of charge on stability of
Reissner-Nordstr$\ddot{o}$m wormholes. Sharif and collaborators
\cite{10}-\cite{10b} explored the role of electric charge on the
stability of spherical as well as cylindrical thin-shell wormholes.
Lobo \emph{et al.} \cite{11} introduced a novel approach to study
wormhole stability under linearized perturbations and analyzed that
matter content supporting wormhole throat minimally violates NEC.

It has been a burning issue for cosmologists to have some physically
suitable models for exotic matter. Different candidates of dark
energy have been proposed in this regard like phantom energy
\cite{15}, quintessence \cite{15a} and family of Chaplygin gas (CG)
\cite{15b}. Eiroa \cite{16} assumed generalized Chaplygin gas (GCG)
to study the dynamics of spherical thin-shell wormholes. Sharif and
Azam \cite{18} discussed stability of Reissner-Nordstr\"{o}m
wormhole configurations by taking modified generalized Chaplygin gas
(MGCG) and found both stable as well as unstable configurations. The
stability of spherically symmetric wormholes have been studied in
the context of two different CG models \cite{19,19aa}.
Mazharimousavi and Halilsoy \cite{19a} analyzed the role of angular
momentum on the stability of counter-rotating wormhole solutions in
the vicinity of linear gas EoS.

This paper investigates stability of regular ABG thin-shell
wormholes by considering different dark energy models. The paper is
organized in the following format. In section \textbf{2}, we
construct regular ABG wormhole solutions and discuss various
physical aspects. The standard approach for stability analysis is
given in section \textbf{3}. Section \textbf{4} deals with stability
formalism of the regular ABG thin-shell wormholes in the context of
linear, logarithmic and CG models. We assume small velocity
dependent perturbations to investigate wormhole stability in section
\textbf{5}. Finally, we provide a brief overview of the obtained
results in the last section.

\section{Regular ABG Thin-Shell Wormhole}

The study on global regularity of BHs has attracted the attention of
many researchers. None of the regular BHs are exact solutions to the
field equations without any physically reasonable source. To derive
the nonlinear electromagnetic field, one requires to enlarge the
class of electrodynamics to nonlinear ones \cite{6b}. These regular
BHs behave as ordinary Reissner-Nordstr\"{o}m BH solutions and the
existence of these solutions does not contradict the singularity
theorems \cite{bb}. This motivates us to discuss stability of viable
wormhole solutions coupled with nonlinear electrodynamics. The
static spherically symmetric nonsingular ABG BH is given by
\begin{equation}\label{1}
ds^2=-G(r)dt^2+G^{-1}(r)dr^2+r^2(d\theta^2+\sin^2\theta d\phi^2),
\end{equation}
where
\begin{equation}\label{1a}
G(r)=1-\frac{2Mr^2}{(r^2+Q^2)^{\frac{3}{2}}}+\frac{r^2Q^2}{(r^2+Q^2)^2},
\end{equation}
$M$ is the mass and $Q$ denotes the charge. We choose this regular
BH for wormhole construction because a regular system can be
constructed from a finite energy and its evolution is more
acceptable. The corresponding spacetime becomes regular at $r = 0$
and behaves as Reissner-Nordstr\"{o}m BH for $r\rightarrow\infty$.

Its event horizon is the largest root of the equation
\begin{equation}\label{2}
r^4+Q^4+3r^2Q^2-2Mr^2\sqrt{r^2+Q^2}=0.
\end{equation}
If real and positive solutions of Eq.(\ref{2}) exist, the spacetime
yields a BH while if there is no such solution then spacetime is
fully regular without event horizon \cite{20a}. The analysis of the
roots shows that the above expression constitutes positive real
roots if $0<Q<2M$. The critical value of charge $Q_{crit}=0.634M$
separates the BH and the "no-horizon" ABG spacetime. In no-horizon
case, the ABG spacetime is regular at all radii $r\geq0$. For $Q\geq
Q_{crit}$, there are no horizons while the given spacetime admits
two horizons for $Q<Q_{crit}$ leading to a non-extremal BH. The
Schwarzschild BH is recovered for $Q=0$. For no-horizon solutions,
ABG spacetime allows circular geodesic motion of test particles and
photons. There exist two photon circular geodesics in the no-horizon
spacetime in which the outer one is unstable relative to radial
perturbations and the inner one is stable \cite{20a,20aa}. We also
consider ABG regular BH coupled with a cosmological constant
($\Lambda$) in the metric function as \cite{20b}
\begin{equation}\label{2a}
G(r)=1-\frac{2Mr^2}{(r^2+Q^2)^{\frac{3}{2}}}+\frac{r^2Q^2}{(r^2+Q^2)^2}
-\frac{\Lambda r^2}{3},
\end{equation}
which reduces to Eq.(\ref{1a}) as $\Lambda\rightarrow0$. This metric
describes a regular ABG BH in de Sitter $(\Lambda>0)$ and anti-de
Sitter $(\Lambda<0)$ spacetimes. The static radius of the given
spacetime plays an important role for the location of its horizons
and no-horizon solutions \cite{20c}-\cite{20ccc}. In the case of de
Sitter and anti-de Sitter regular ABG spacetimes, the metric
function $G(r)$ can have at most three distinct roots leading to
three horizons in which $r_{c}$ is the cosmological horizon, $r_{h}$
is the event horizon and $r_{b}$ is the inner horizon \cite{20d}.

For the construction of a timelike thin-shell wormhole, we apply
standard cut and paste method such that the resulting 4D copies are
glued at the hypersurface $\Sigma^\pm=\Sigma=\{r=a\}$. The wormhole
throat should fulfill the flare-out condition for which exotic
matter is required. To avoid singularities and unification of event
and cosmological horizons, we must have $r_{h}<a_{0}<r_{c}$ for the
possibility of static wormhole solutions from de Sitter and anti-de
Sitter spacetimes. The existence of horizons in wormhole
configuration do not matter because the wormhole throat must be
taken out of horizon, i.e., $a_{0}>r_{h}$ for traversability of
thin-shell wormholes \cite{20f}. We consider a 3D timelike spacetime
at the shell defined by
\begin{equation}\label{4}
ds^2=-d\tau^2+a^2(\tau)(d\theta^2+\sin^2\theta d\phi^2),
\end{equation}
where $\tau$ is the proper time on the shell. The unit four-vector
normals $n^\pm_{\alpha}$ to $\mathcal{M}^\pm$ are defined by
\begin{equation}\label{7}
n^\pm_{\alpha}=\pm\left|g^{\mu\nu}\frac{\partial\eta}{\partial
x^\mu}\frac{\partial\eta}{\partial
x^\nu}\right|^{-\frac{1}{2}}\frac{\partial\eta}{\partial
x^\alpha}=\left(-\dot{a},\frac{\sqrt{G(r)+\dot{a}^2}}{G(r)},0,0\right),
\end{equation}
satisfying the relation $n^\alpha n_{\alpha}=\epsilon=1$. We
implement Israel formalism for the dynamical evolution of thin-shell
which allows matching of two spacetime regions separated by
$\Sigma$. We use Lanczos equations to determine surface stresses at
wormhole throat given by
\begin{equation}\label{5}
S^i_{j}=\frac{1}{8\pi}\left\{[K]\delta^i_{j}-[K^i_{j}]\right\},
\end{equation}
where $[K^i_{j}]$ represents the extrinsic curvature tensor and
$K=tr[K^i_{j}]$. The surface stresses at the shell are determined as
\cite{10}
\begin{eqnarray}\label{9}
\sigma&=&-\frac{1}{2\pi a}\sqrt{G(a)+\dot{a}^2},\\\label{10}
p&=&\frac{1}{4\pi}\left[\frac{\sqrt{G(a)+\dot{a}^2}}{a}
+\frac{2\ddot{a}+G'(a)}{2\sqrt{G(a)+\dot{a}^2}}\right].
\end{eqnarray}
For the existence of any physically realistic matter, we need to
check the energy conditions. The sum of surface stresses of matter
indicates violation of NEC leading to the presence of exotic matter.
The amount of this matter should be minimized for the sake of viable
wormhole configurations. We see from Figure \textbf{1} that
$\sigma<0$ and $\sigma+p<0$ showing the violation of NEC and WEC for
different values of $Q$ and $a$ (with and without $\Lambda$).
\begin{figure}\center
\epsfig{file=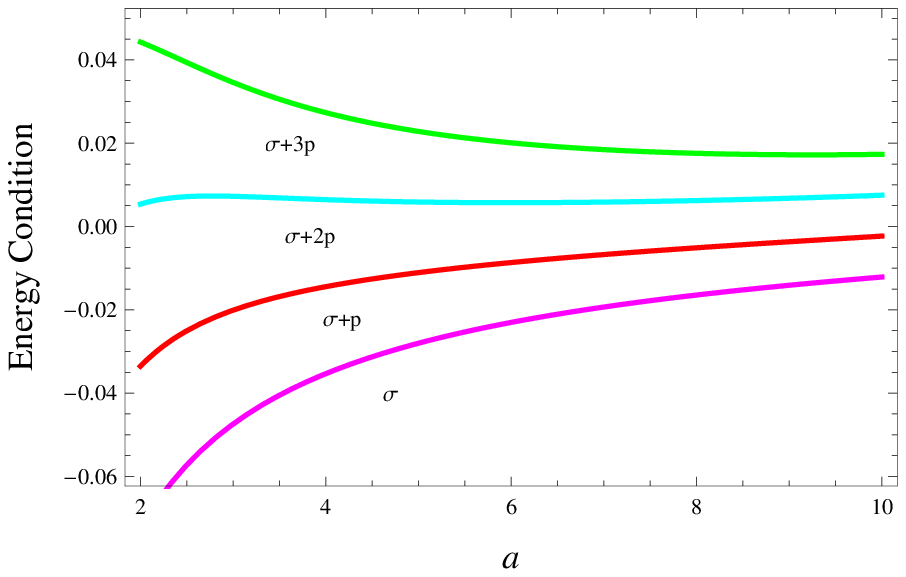,width=0.49\linewidth}\epsfig{file=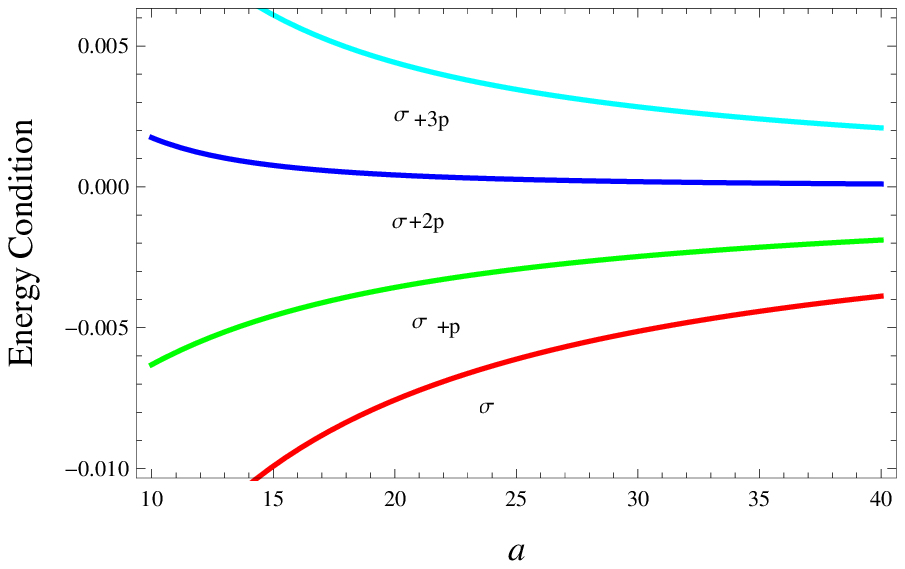,width=0.49\linewidth}\\
\caption{Plots of energy conditions with $\Lambda=0.1$ (left hand
side) and without $\Lambda$ (right hand side) corresponding to
$\frac{Q}{M}=1.1$.}
\end{figure}

Now we analyze attractive and repulsive behavior \cite{21} of the
constructed wormhole solutions for which observer's
four-acceleration is given by
\begin{equation}\nonumber
a^\mu=u_{;\nu}^\mu u^\nu,
\end{equation}
where $u^\mu=\frac{dx^\mu}{d\tau}=(\frac{1}{\sqrt{G(r)}},0,0,0)$ is
the observer's four-velocity. The non-zero four-acceleration
component is computed as
\begin{equation}\label{12a}
a^r=\Gamma^r_{tt}\left(\frac{dt}{d\tau}\right)^2=\frac{3Mr^3}
{(r^2+Q^2)^{\frac{5}{2}}}-\frac{2Mr}
{(r^2+Q^2)^{\frac{3}{2}}}+\frac{rQ^2}{(r^2+Q^2)^2}-\frac{2r^3Q^2}
{(r^2+Q^2)^3}-\frac{\Lambda r}{3},
\end{equation}
and the geodesic equation has the following form
\begin{equation}\nonumber
\frac{d^2r}{d\tau^2}=-\Gamma^r_{tt}\left(\frac{dt}{d\tau}\right)^2=-a^r.
\end{equation}
A wormhole will be attractive in nature if $a^r>0$ while it will
exhibit repulsive characteristics for $a^r<0$. The attractive and
repulsive characteristics of the regular ABG thin-shell wormholes
are analyzed corresponding to different values of charge. Figure
\textbf{2} shows respective plots with and without $\Lambda$.
\begin{figure}\center
\epsfig{file=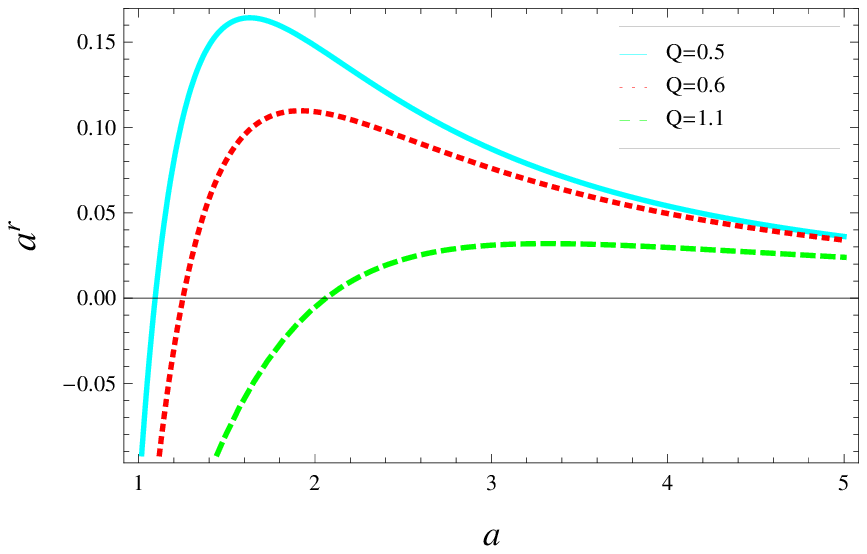,width=0.5\linewidth}\epsfig{file=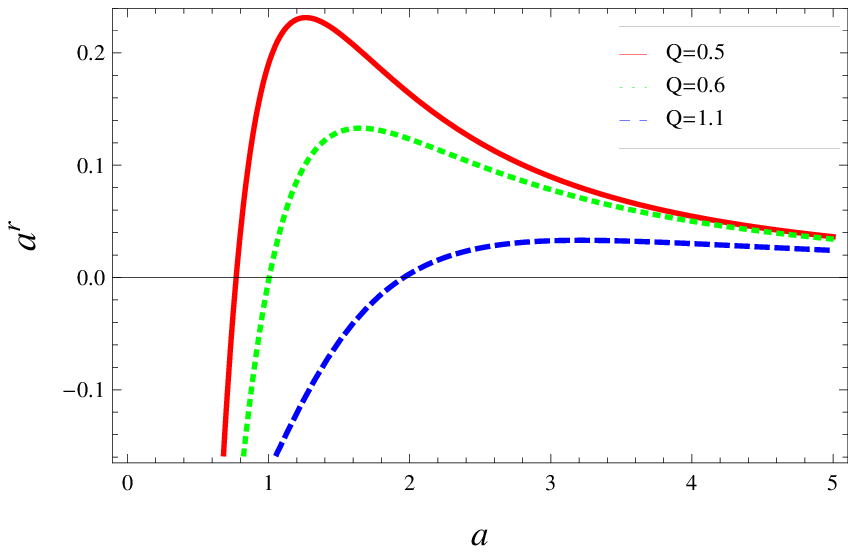,width=0.5\linewidth}\\
\caption{Plots of $a^r$ with $\Lambda=0.1$ (left hand side) and
without $\Lambda$ (right hand side) for different values of
$\frac{Q}{M}$. The wormhole will be attractive or repulsive
corresponding to $a^r>0$ or $a^r<0$, respectively.}
\end{figure}

It is well-known fact that exotic matter supports the wormhole
throat to make it traversable. The total amount of this matter at
shell is quantified by the integral theorem \cite{5}
\begin{equation}
\Omega=\int^{2\pi}_{0}\int^{+\infty}_{-\infty}[\rho+p_{r}]\sqrt{-g}
dRd\phi.
\end{equation}
The wormhole shell, being thin, does not apply any pressure
($p_{r}=0$). Taking $\rho=\delta(R)\sigma(a)$, we have
\begin{equation}\label{12b}
\Omega_{a}=\int^{2\pi}_{0}[\rho\sqrt{-g}]|_{r=a}d\phi=2\pi
a\sigma(a).
\end{equation}
This expression through surface energy density $\sigma(a_{0})$
becomes
\begin{equation}
\Omega_{a}=-\frac{\left(a_{0}^4+3a_{0}^2Q^2-2Ma_{0}^2\sqrt{a_{0}^2+Q^2}+
Q^4-\frac{\Lambda
a_{0}^2}{3}(a_{0}^2+Q^2)^2\right)^{\frac{1}{2}}}{a_{0}^2+Q^2},
\end{equation}
where $a_{0}$ represents static throat radius of ABG thin-shell
wormhole. The behavior of exotic matter is given in Figure
\textbf{3} which shows how the amount of exotic matter varies by
varying charge.
\begin{figure}\center
\epsfig{file=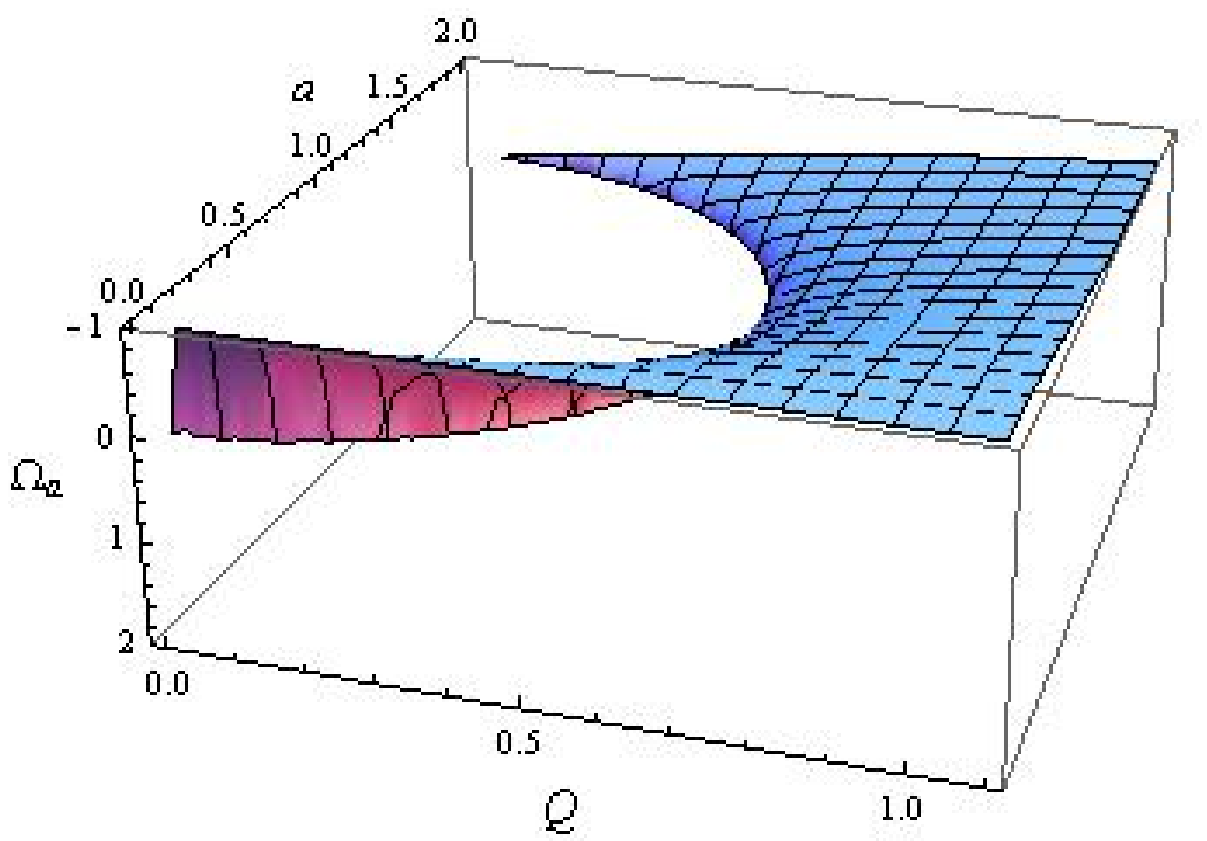,width=0.53\linewidth}\epsfig{file=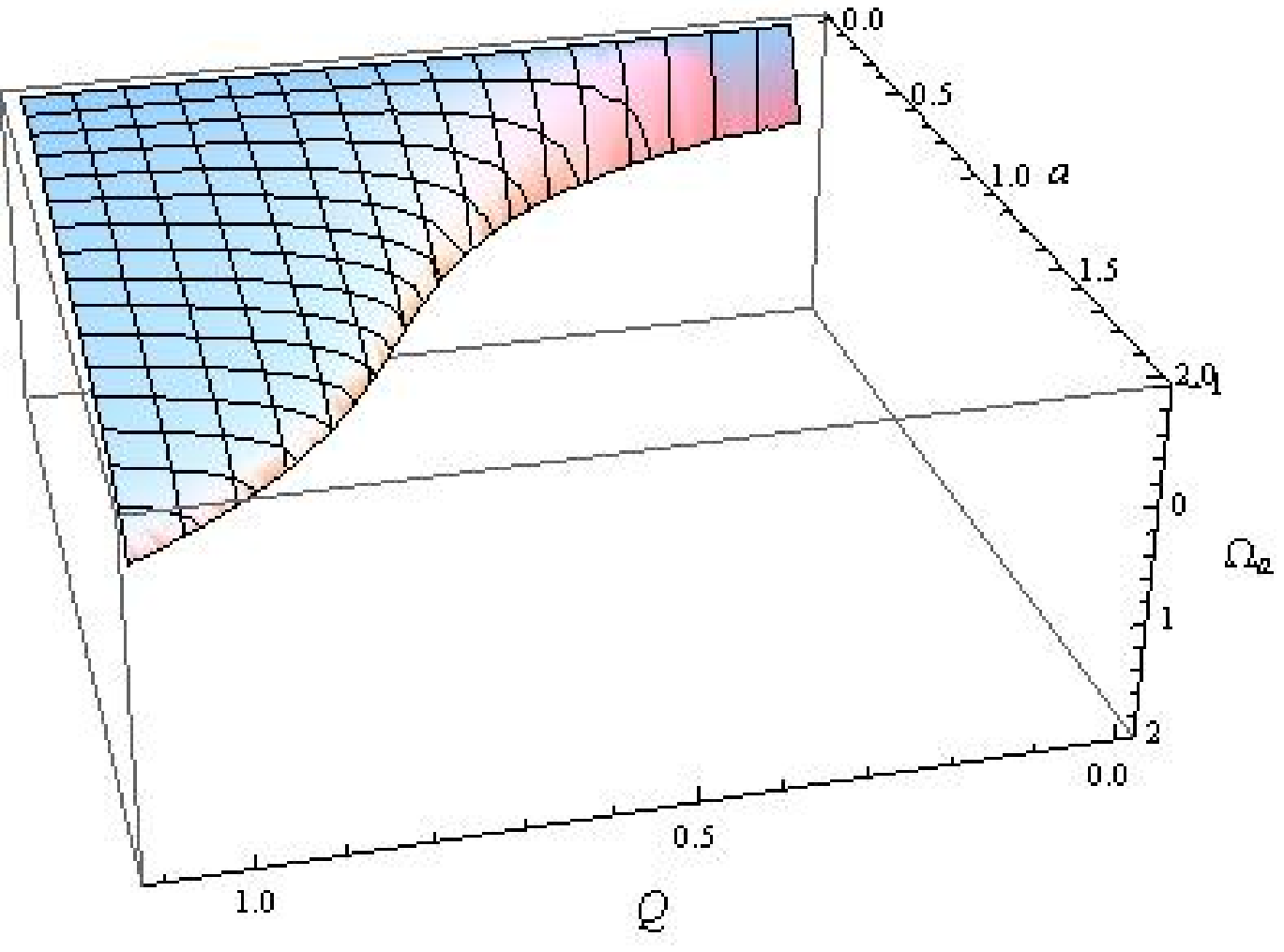,width=0.49\linewidth}\\
\caption{Plots for the total amount of exotic matter with
$\Lambda=0.1$ (left hand side) and without $\Lambda$ (right hand
side) with different values of charge.}
\end{figure}

\section{Conditions for Wormhole Stability }

In this section, we provide a standard approach for the wormhole stability under
linear perturbations. The surface stresses corresponding to static
wormhole solution ($a=a_{0}$) become
\begin{equation}\label{12}
\sigma_{0}=-\frac{\sqrt{G(a_{0})}}{2\pi a_{0}}, \quad
p_{0}=\frac{1}{4\pi}\left[\frac{\sqrt{G(a_{0})+}}{a_{0}}
+\frac{G'(a_{0})}{2\sqrt{G(a_{0})}}\right].
\end{equation}
In order to explore wormhole stability, we consider barotropic EoS as a linear
perturbation in the form
\begin{equation}\label{12a}
p=\Phi(\sigma),
\end{equation}
where $\Phi(\sigma)$ is chosen arbitrarily. This
covers the polytropic EoS $p\approx\sigma^{1+\frac{1}{n}}$ with
$0\leq n<\infty$. The energy density and pressure follow the
conservation identity $S_{~;j}^{ij}=0$ which, for the induced metric
(\ref{4}), becomes
\begin{equation}\label{13}
\frac{d}{d\tau}(\sigma\Psi)+p\frac{d\Psi}{d\tau}=0,
\end{equation}
where $\Psi=4\pi a^2$ is the area of wormhole throat. The equation
of motion for thin-shell can be obtained by setting Eq.(\ref{9}) as
$\dot{a}^2+\Delta(a)=0$, which describes wormhole dynamics whereas
$\Delta(a)$ is the potential function defined by
\begin{equation}\label{16}
\Delta(a)=G(a)-[2\pi a\sigma]^2.
\end{equation}
The stability of wormhole static solution requires
$\Delta'(a_{0})=0=\Delta(a_{0})$ and $\Delta''(a_{0})>0$. In this
context, we substitute Eq.(\ref{12a}) and
$\sigma'=\frac{\dot{\sigma}}{\dot{a}}$ in conservation equation
which yields
\begin{equation}\label{17}
\sigma'=-\frac{2}{a}(\sigma+\Phi),
\end{equation}
leading to
\begin{equation}\label{18}
\sigma''=\frac{2}{a^2}(\sigma+\Phi)(3-a\Phi').
\end{equation}
The first derivative of Eq.(\ref{16}) through (\ref{17}) takes the
form
\begin{equation}\label{20}
\Delta'(a_{0})=G'(a_{0})+8\pi^2
a_{0}\sigma_{0}[\sigma_{0}+2p(\sigma_{0})],
\end{equation}
which further turns out to be
\begin{eqnarray}\label{21a}
\Delta''(a_{0})&=&G''(a_{0})-8\pi^2\left\{[\sigma_{0}+2p_{0}]^2+2
\sigma_{0}[\sigma_{0}+p_{0}][1+2\Phi'(\sigma_{0})]\right\},
\end{eqnarray}
where $\Phi_{0}=p_{0}$.

\section{Models for Exotic Matter}

Here we study stability of regular ABG thin-shell wormholes in the
context of different dark energy models. The choice of model has
remarkable significance in the dynamical investigation of thin-shell
wormholes. Recently, the dynamics of regular wormhole solutions has
been examined by taking linear, logarithmic and Chaplygin gas models
\cite{6d,21a} . In this paper, we consider these fluids to study
stability of regular ABG wormholes. This enables us to examine the
effect of electric charge and other parameters on the wormhole
stability. In the following, we discuss stability analysis in the
context of above mentioned candidates for exotic matter.

\subsection{Linear Gas}

We consider a linear gas \cite{21} defined by an EoS
\begin{equation}\label{11}
\Phi=p_{0}+\mu(\sigma-\sigma_{0}),
\end{equation}
here $\mu$ represents a constant parameter. The first derivative of
this equation w.r.t. $\sigma$ leads to $\Phi'(\sigma_{0})=\mu$. It
is found that $\Delta(a)$ and $\Delta'(a)$ vanish by substituting
$\sigma(a_{0})$ and $p(a_{0})$. We explore the possibility for the
existence of stable wormhole solutions with increasing values of
electric charge. Figure \textbf{4} shows stable solutions (red
curves) of the regular ABG wormholes with $\frac{Q}{M}=0, ~0.634,
~0.77, ~1.1$. Here $\frac{Q}{M}=0$ represents the Schwarzschild
case. We also plot the metric function $G(r)$ to evaluate the
position of wormhole throat and event horizon of ABG BH, where we
choose $a_{0}=r$ at thin-shell (hypersurface $\Sigma$). It is
mentioned here that we choose the same range for both $a_{0}$ and
$r-$values to plot the metric function $G(r)$. To avoid the
existence of event horizon in the wormhole configuration, we take
$a>r_{h}$ for traversable thin-shell wormholes. We analyze more
stable solutions for positive as well as negative values of
parameter $\mu$ with $\frac{Q}{M}=0.634$. It is observed that the
stability region decreases by increasing the values of
$\frac{Q}{M}$.

We also analyze the stability of regular ABG wormhole configurations
in de Sitter $(\Lambda=0.1)$ and anti-de Sitter $(\Lambda=-0.5)$
backgrounds. The wormhole throat must satisfy $r_{h}<a_{0}<r_{c}$
for the viability of wormhole solutions. The respective results show
more stable wormholes for de Sitter case as compared to anti-de
Sitter spacetime (Figure \textbf{5}). This supports the fact that
effects of linear gas become more significant for the regions of
stability in de Sitter background.
\begin{figure}\center
\epsfig{file=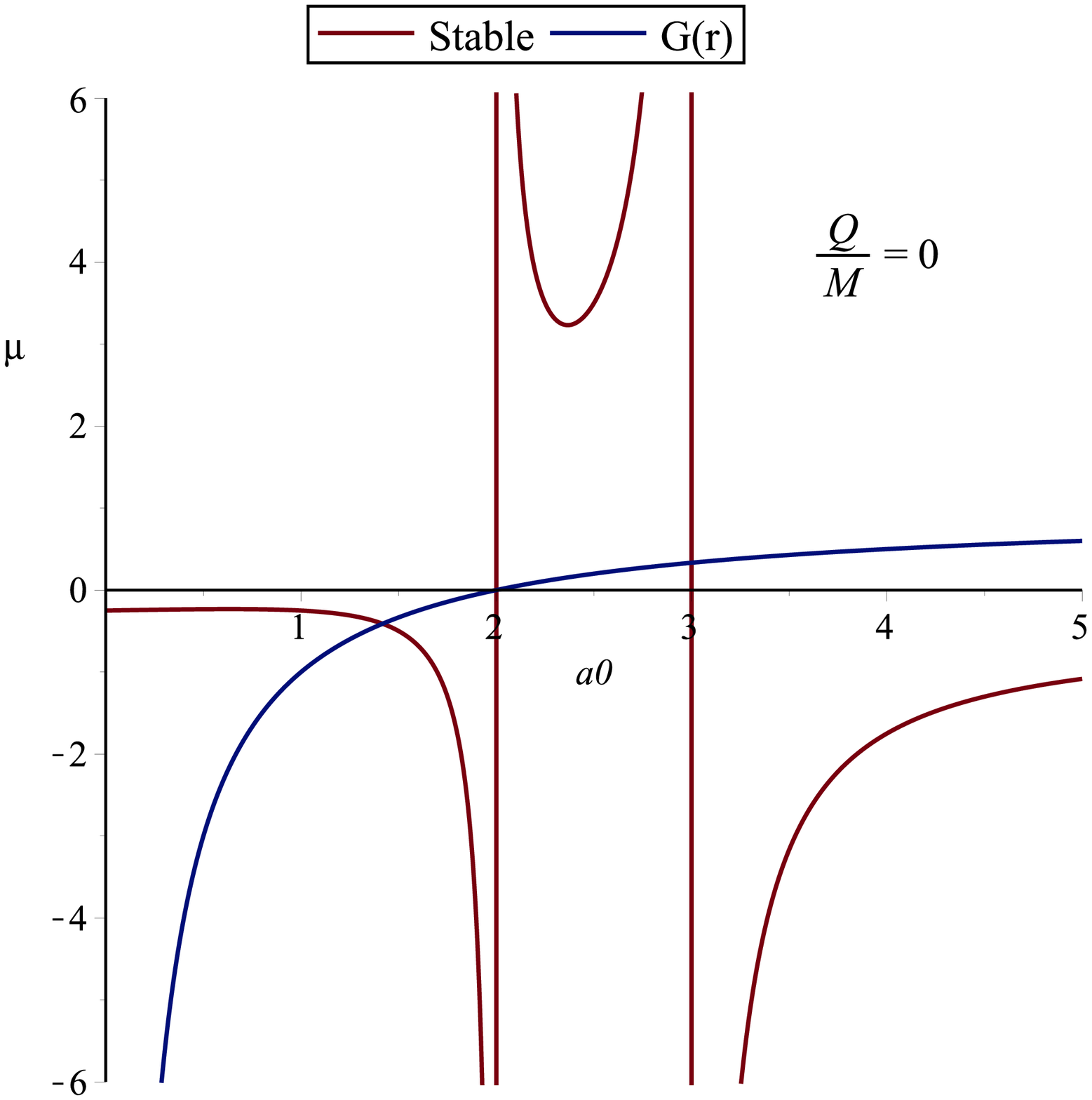,width=0.45\linewidth}\epsfig{file=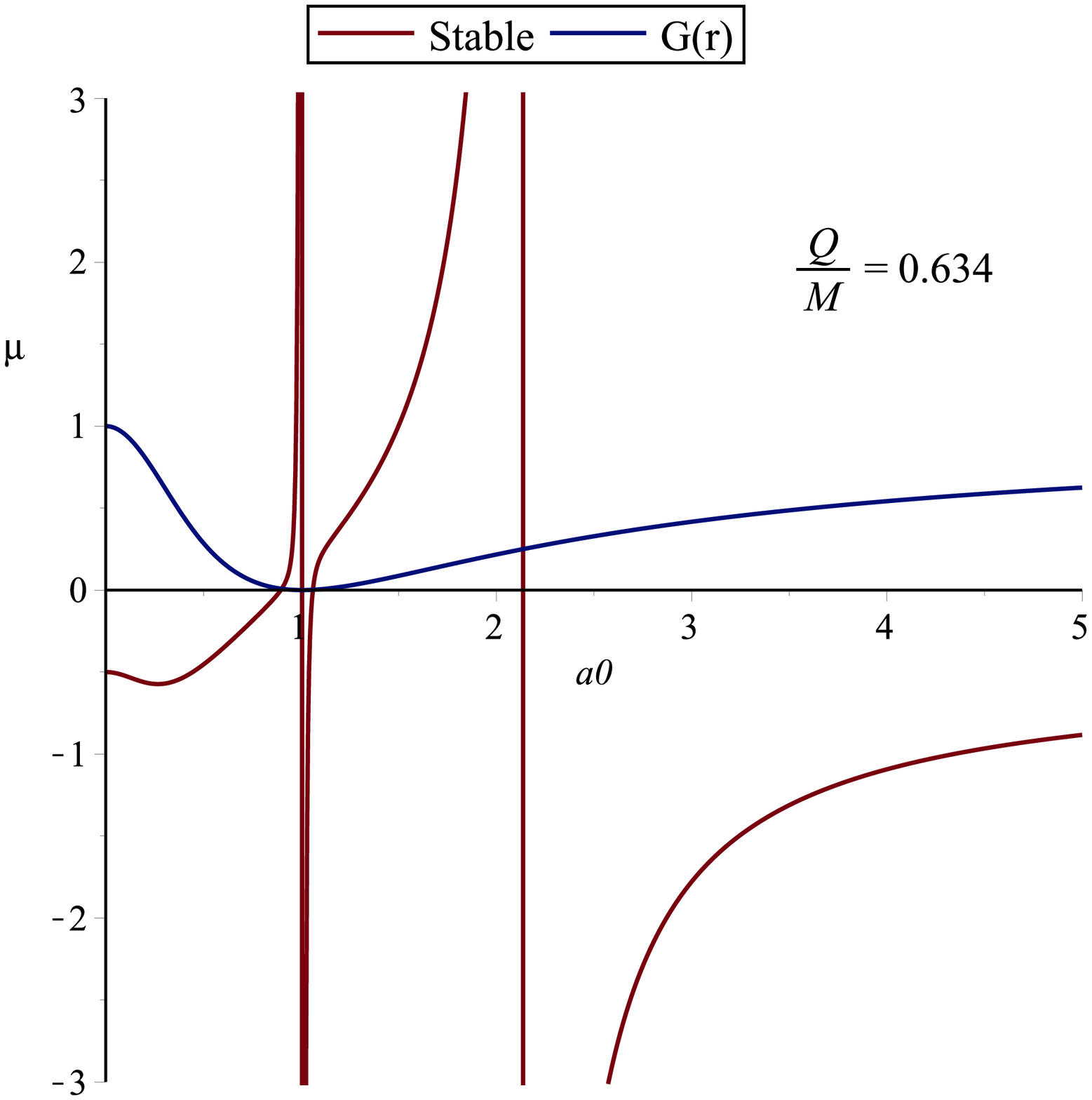,
width=0.45\linewidth}\\
\epsfig{file=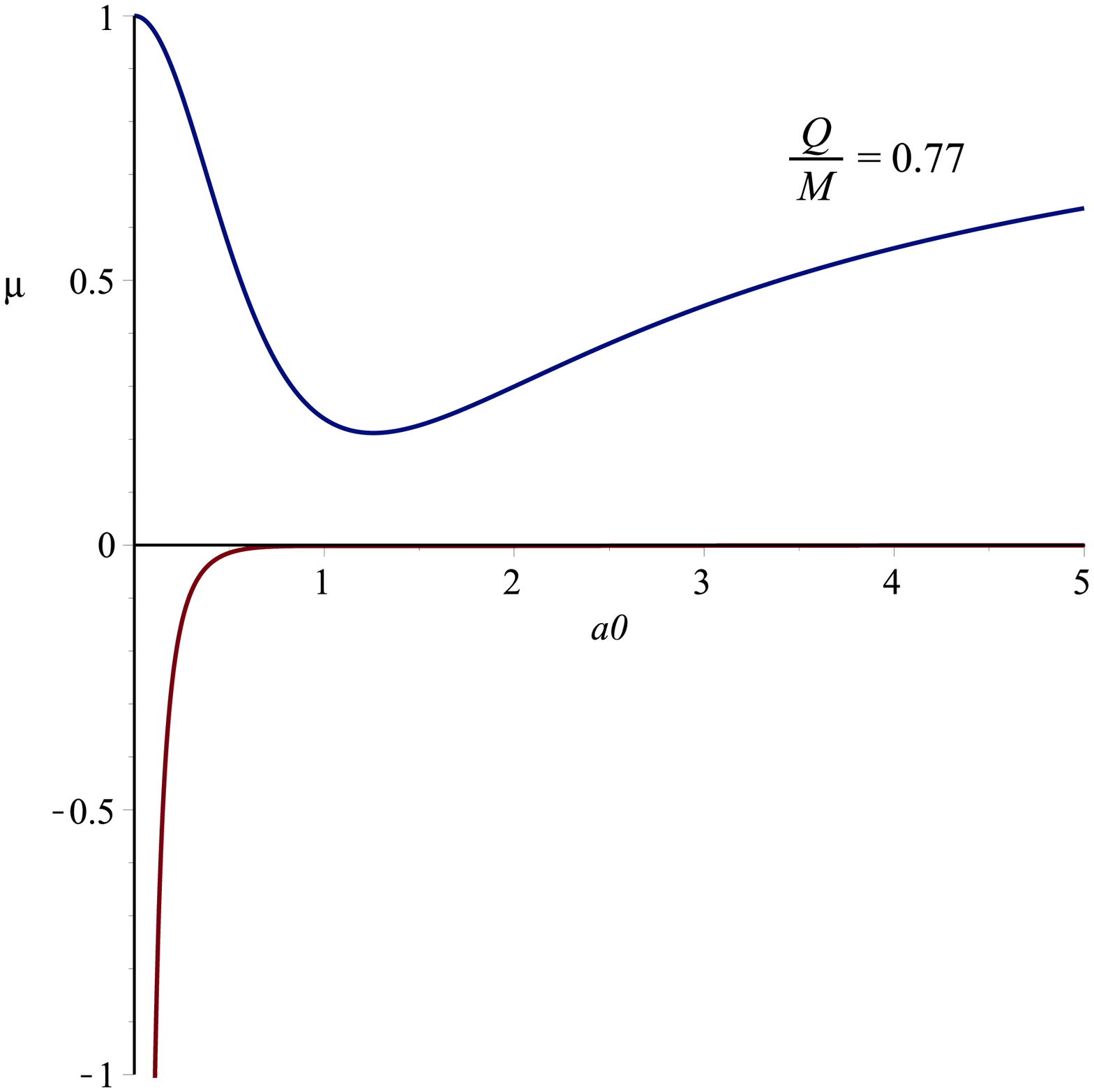,width=0.45\linewidth}\epsfig{file=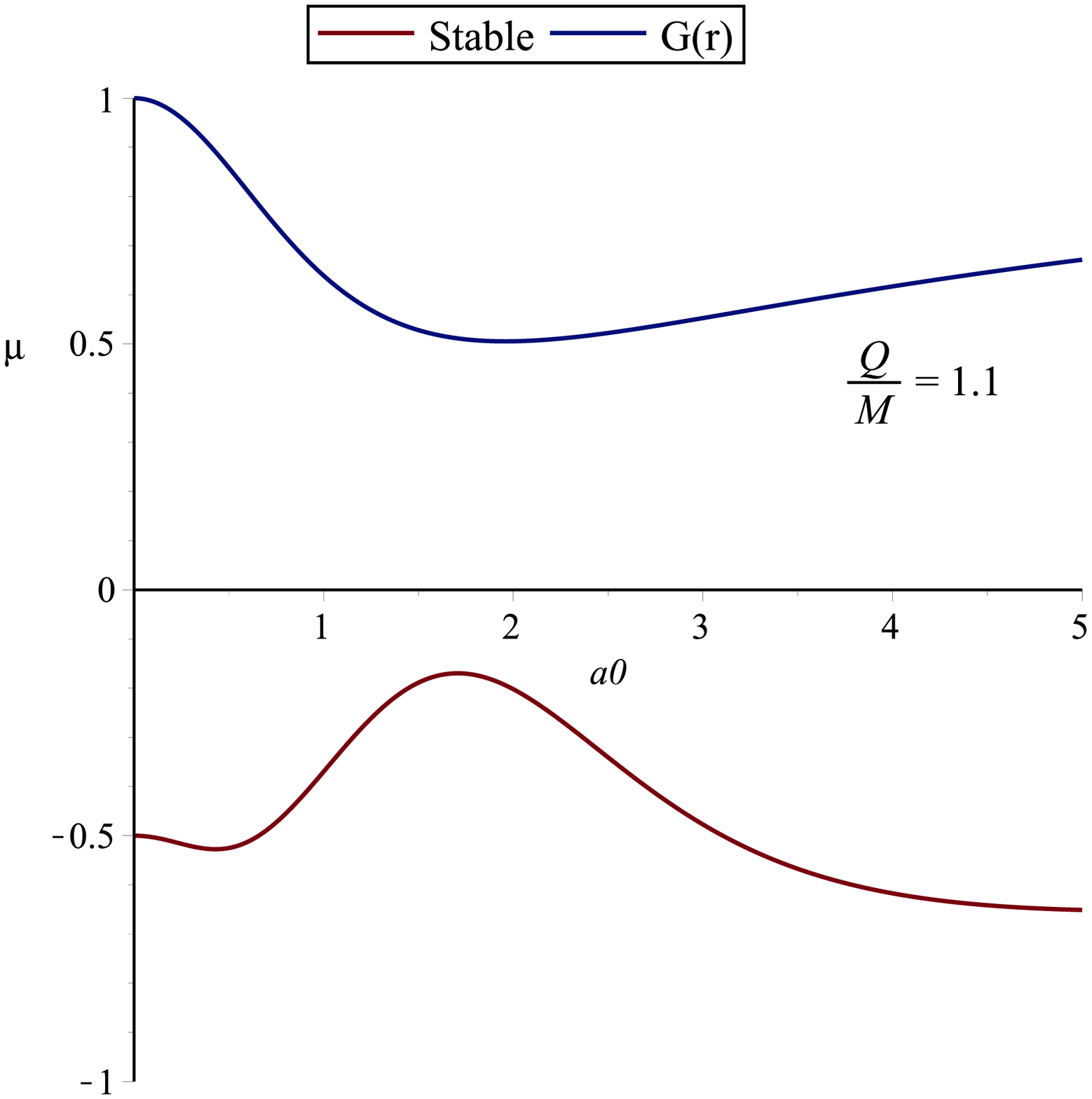,
width=0.45\linewidth}\caption{Plots for regular ABG wormholes by
taking linear gas EoS and $\frac{Q}{M}=0, ~0.634, ~0.77, ~1.1$. The
red and blue curves corresponds to stable regions and the metric
function, respectively. We plot throat radius $a_{0}$ and parameter
$\mu$ along abscissa and ordinate, respectively, where $a_{0}=r$ at
thin-shell.}
\end{figure}
\begin{figure}\center
\epsfig{file=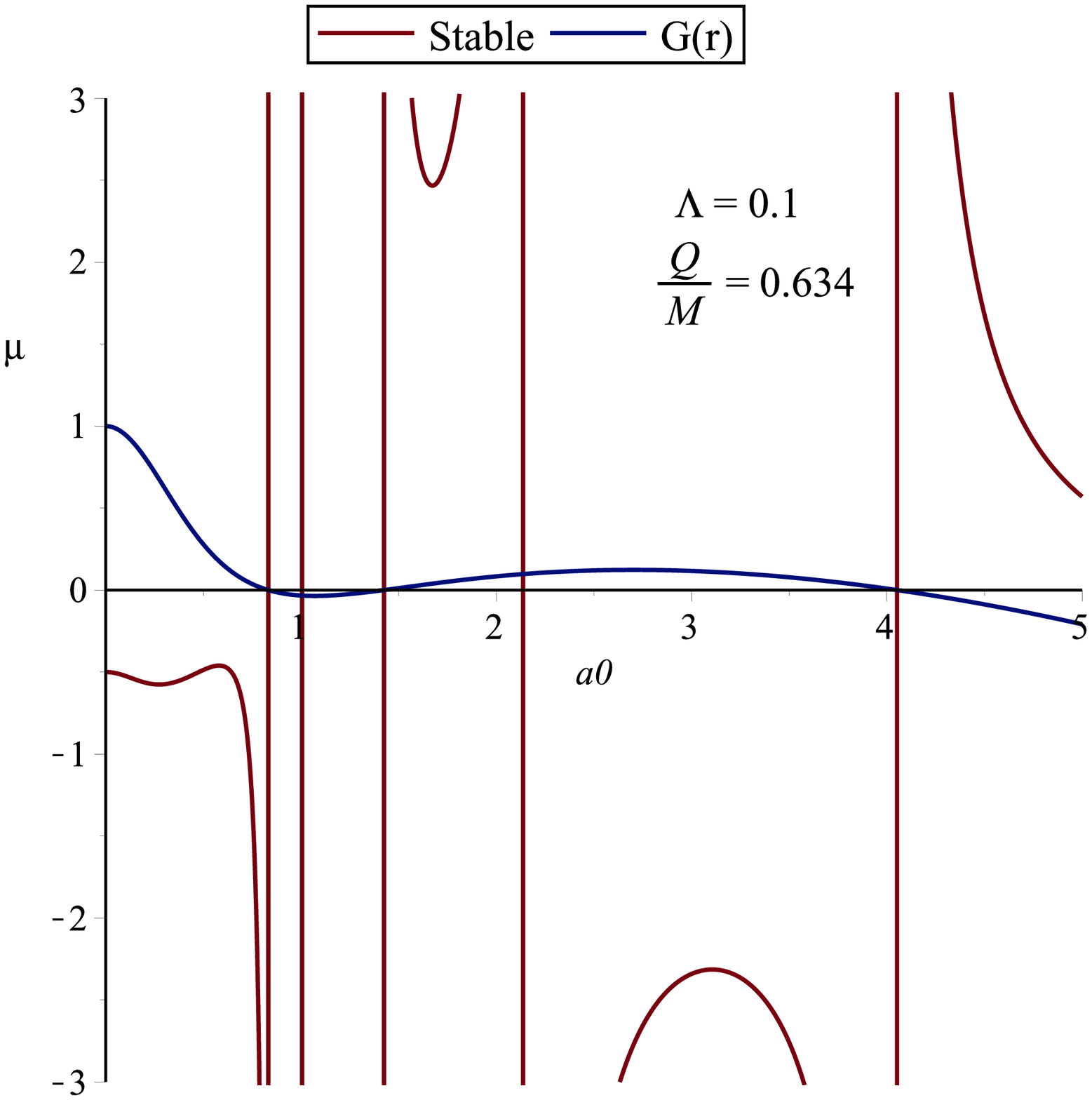,width=0.45\linewidth}\epsfig{file=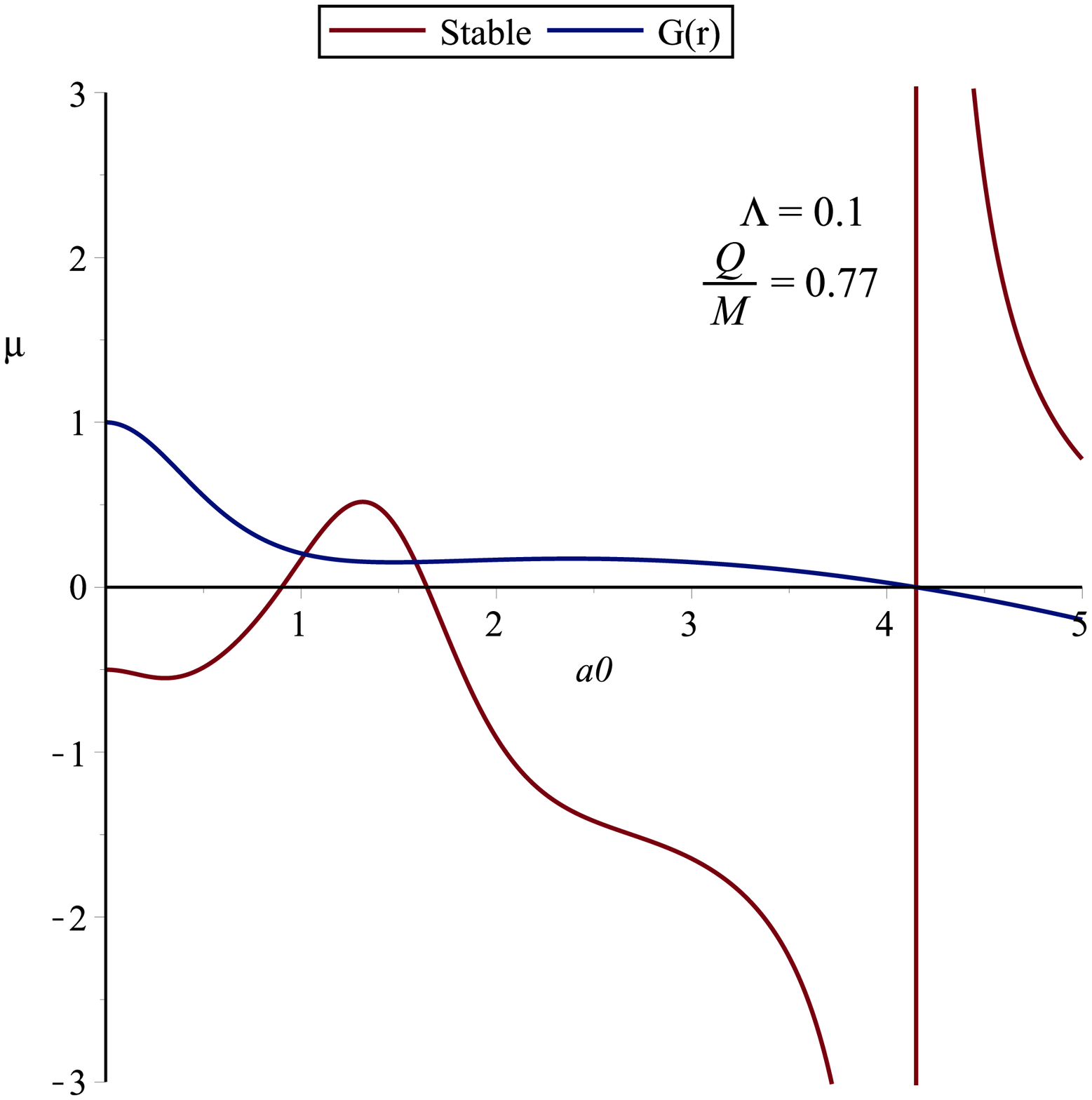,
width=0.45\linewidth}\\
\epsfig{file=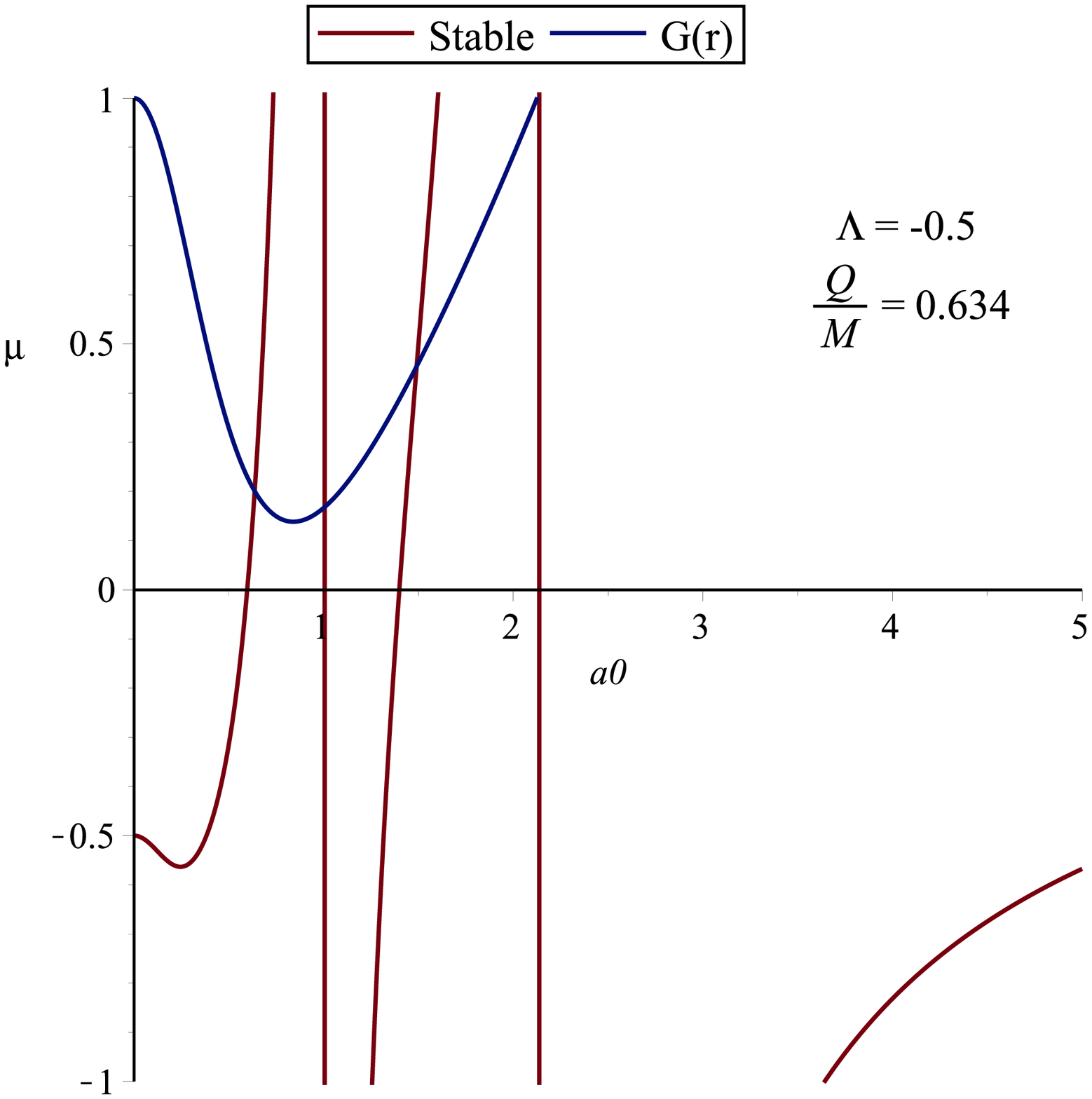,width=0.45\linewidth}\epsfig{file=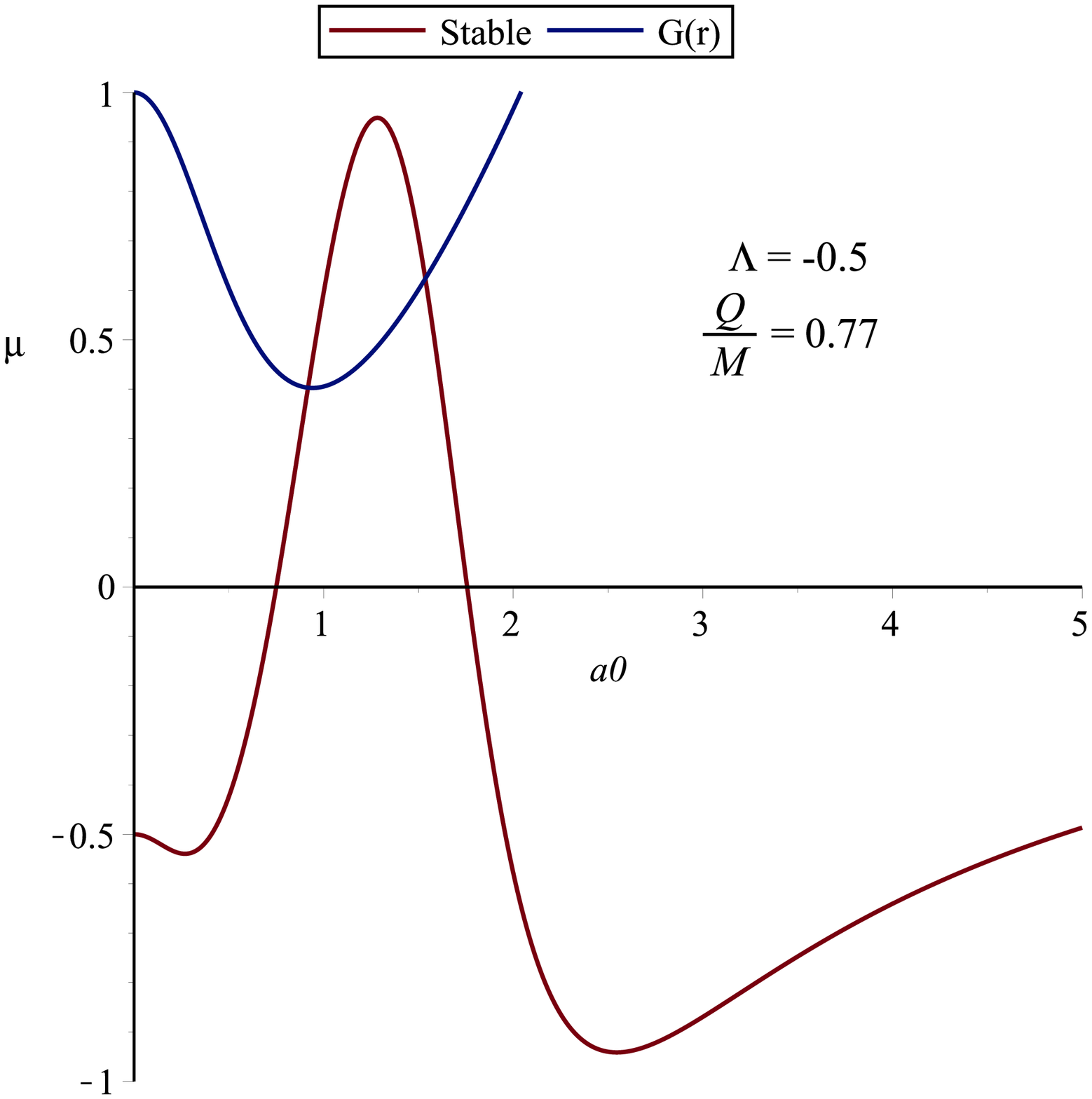,
width=0.45\linewidth}\caption{Plots of stable regular ABG wormholes
with linear gas in de Sitter ($\Lambda=0.1$) and anti-de Sitter
($\Lambda=-0.5$) backgrounds.}
\end{figure}

\subsection{Chaplygin Gas}
\begin{figure}\center
\epsfig{file=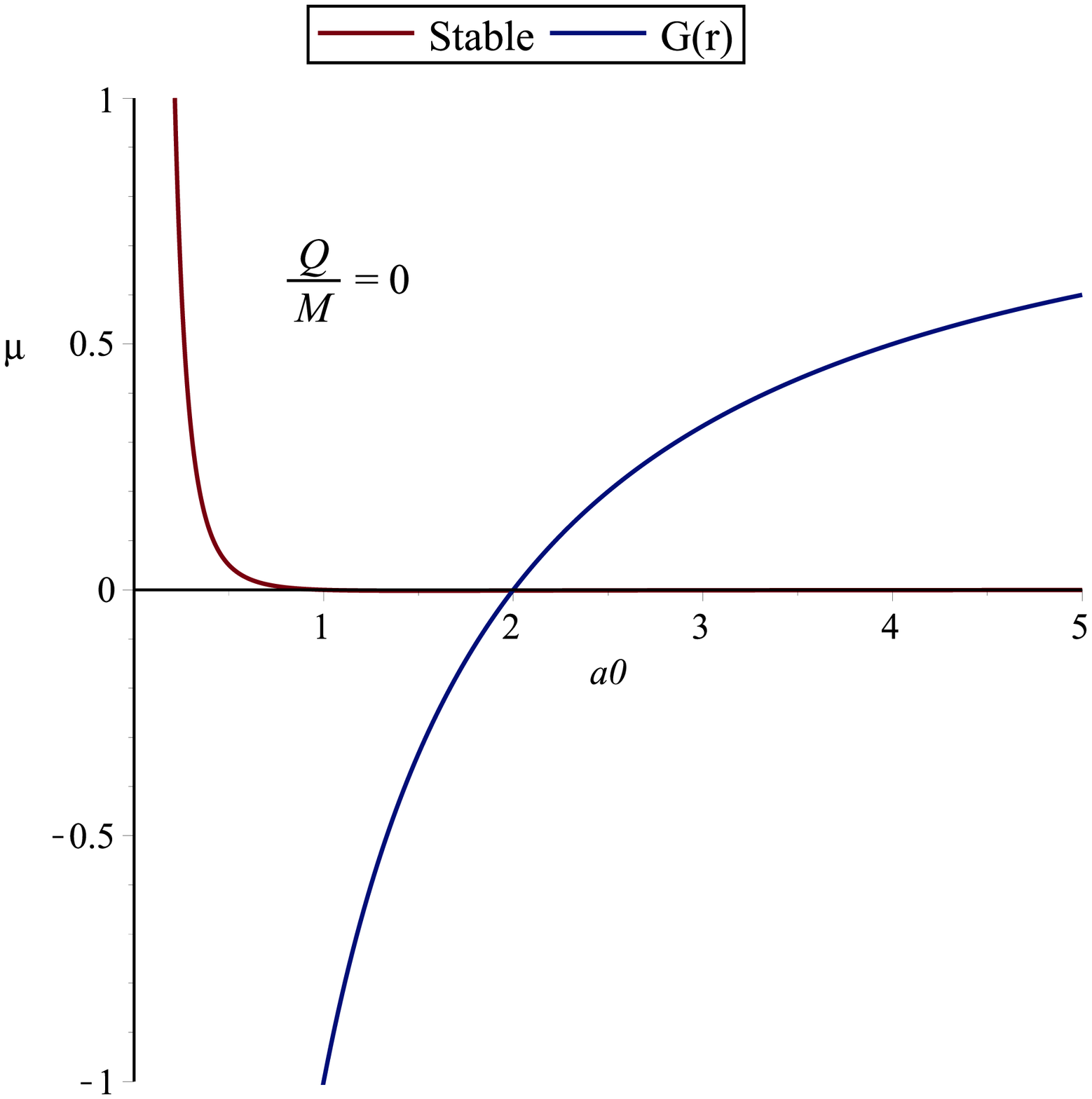,width=0.45\linewidth}\epsfig{file=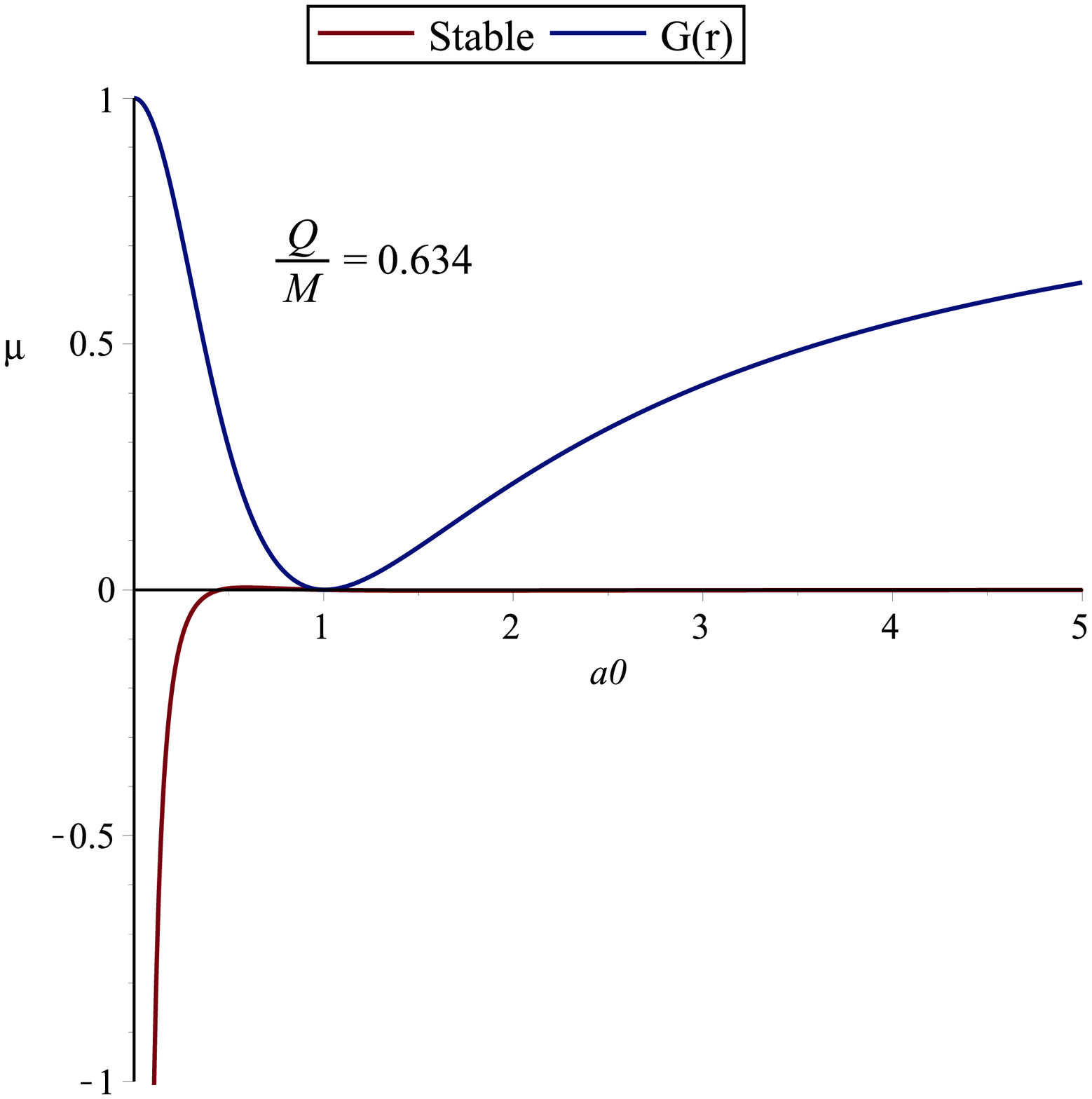,
width=0.45\linewidth}\\
\epsfig{file=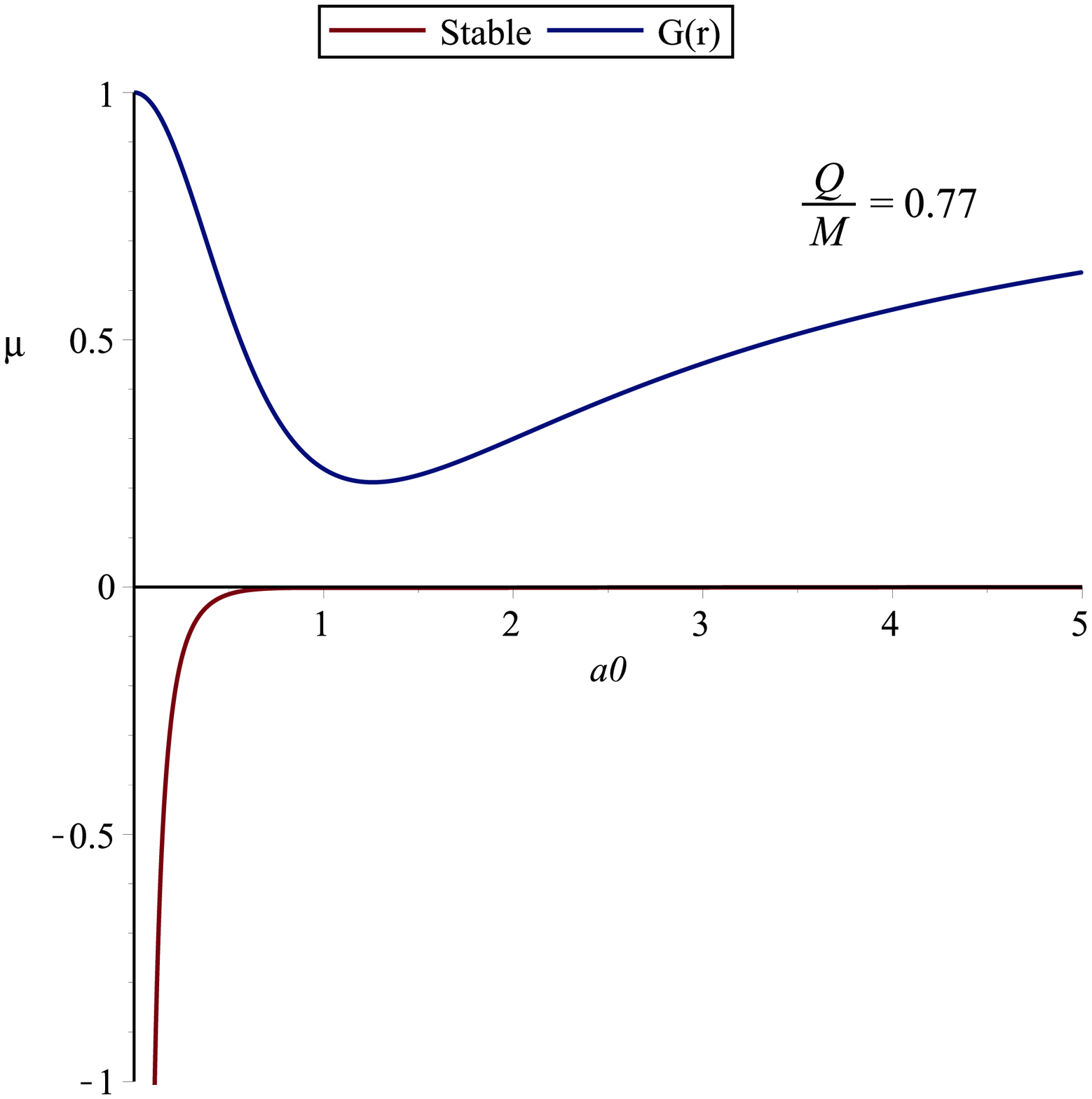,width=0.45\linewidth}\epsfig{file=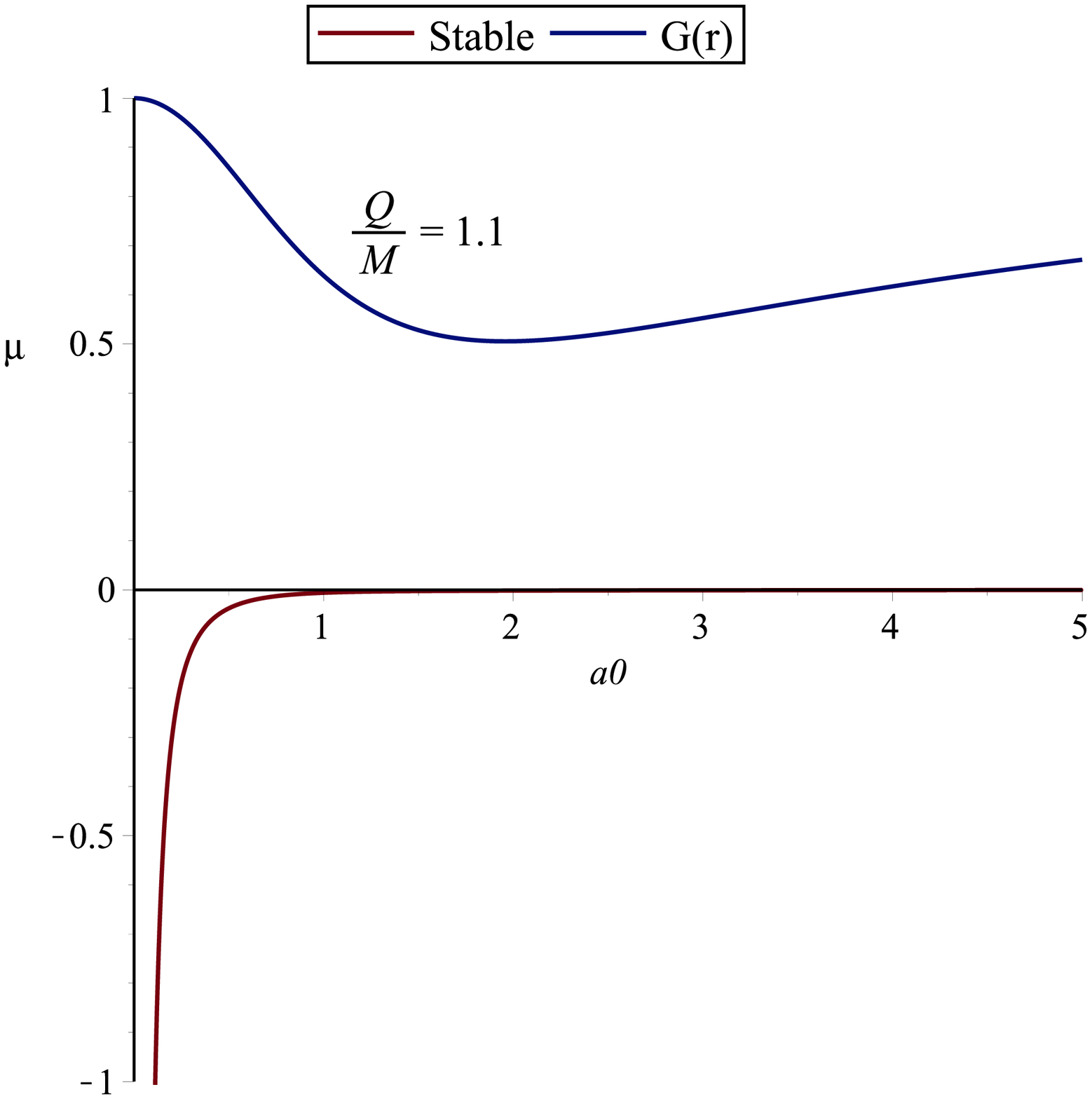,
width=0.45\linewidth}\caption{Plots for stable ABG wormhole
solutions with CG EoS and $\frac{Q}{M}=0, ~0.634, ~0.77, ~1.1$. We
plot throat radius $a_{0}$ and parameter $\mu$ along abscissa and
ordinate, respectively, where $a_{0}=r$ at thin-shell.}
\end{figure}
\begin{figure}\center
\epsfig{file=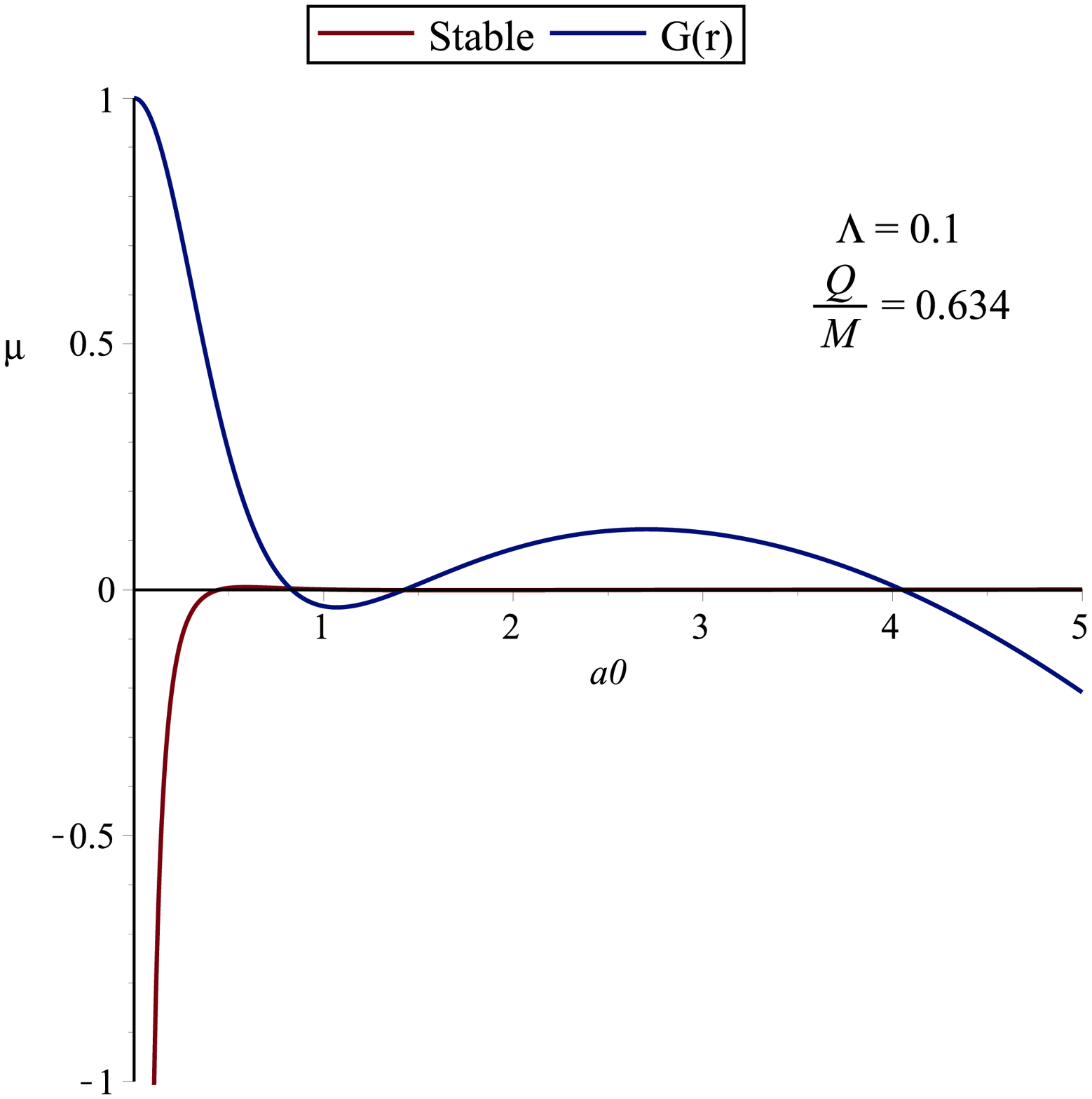,width=0.45\linewidth}\epsfig{file=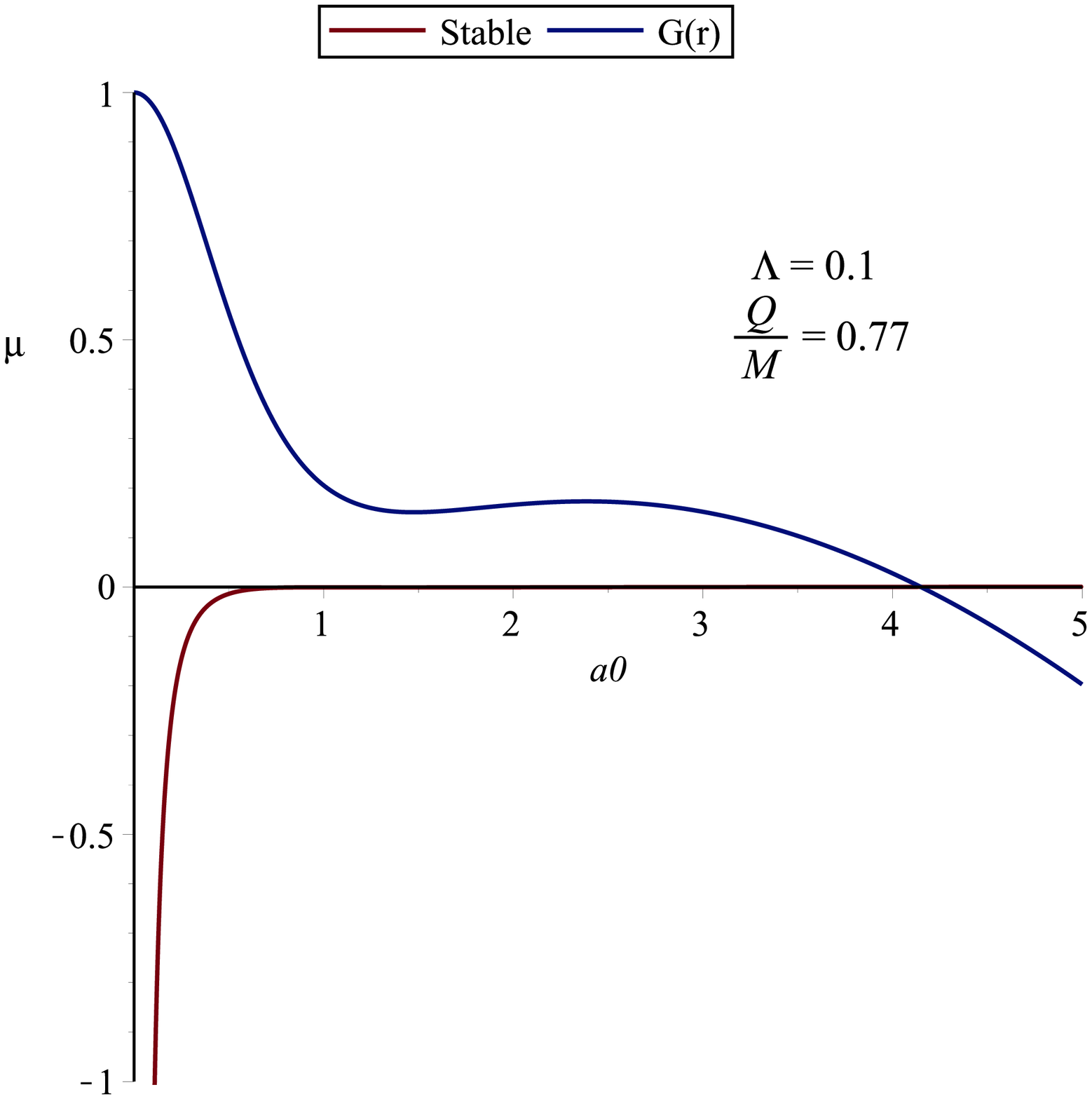,
width=0.45\linewidth}\\
\epsfig{file=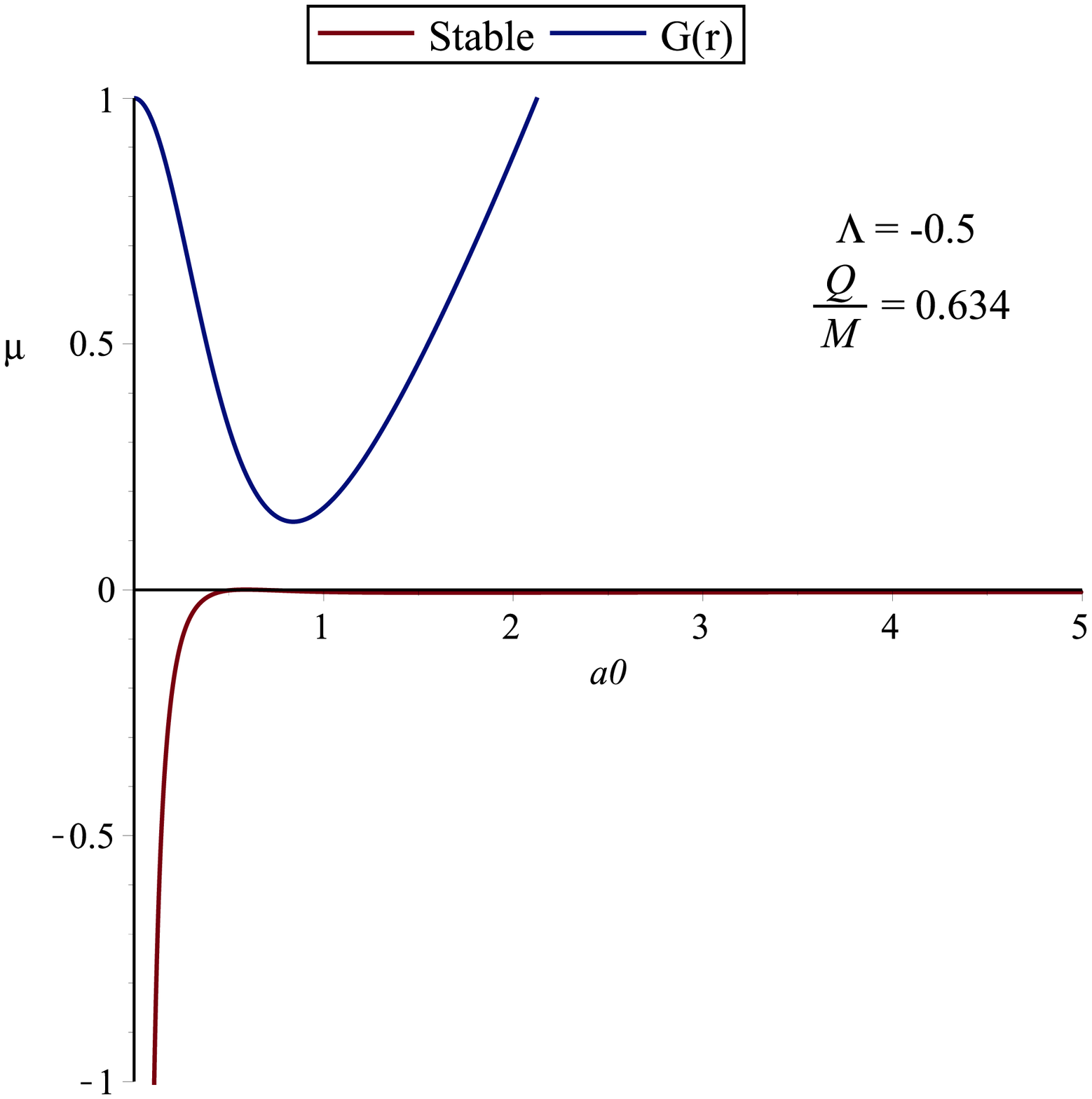,width=0.45\linewidth}\epsfig{file=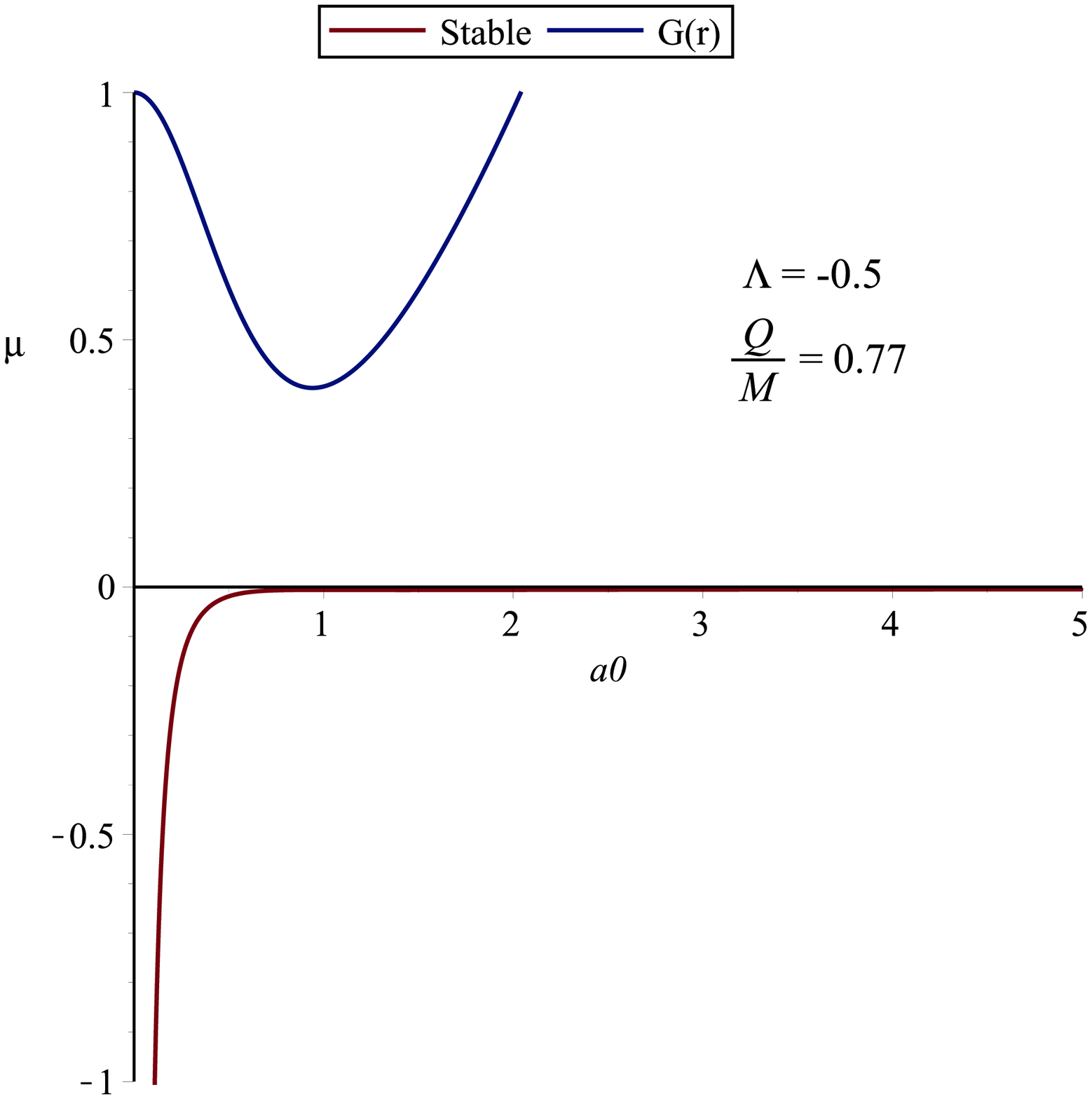,
width=0.45\linewidth}\caption{Plots for stable ABG wormholes with CG
EoS, $\Lambda=0.1, -0.5$ and $\frac{Q}{M}=0.634, ~0.77$.}
\end{figure}

Now we assume CG governed by an EoS of the form
\begin{equation}\label{11a}
\Phi(\sigma)=p_{0}+\mu\left(\frac{1}{\sigma}-\frac{1}
{\sigma_{0}}\right),
\end{equation}
where $\Phi'(\sigma_{0})=-\frac{\mu}{\sigma_{0}^2}$. The results in
Figures \textbf{6} correspond to CG with different values of charge.
We also plot the graphical results for de Sitter and anti-de Sitter
spacetimes as shown in Figure \textbf{7}. In all these cases, only
one stable solution is investigated for different values of
$\frac{Q}{M}$. We find one stable region for $\frac{Q}{M}=0.77,
~1.1$ with negative values of $\mu$. It is found that CG gives
minimum stable regions for the respective wormhole solutions as
compared to the other EoS for exotic matter. We also analyze that
negative values of $\mu$ provide one stable solution in de Sitter
and anti-de Sitter cases.

\subsection{Generalized Chaplygin Gas}
\begin{figure}\center
\epsfig{file=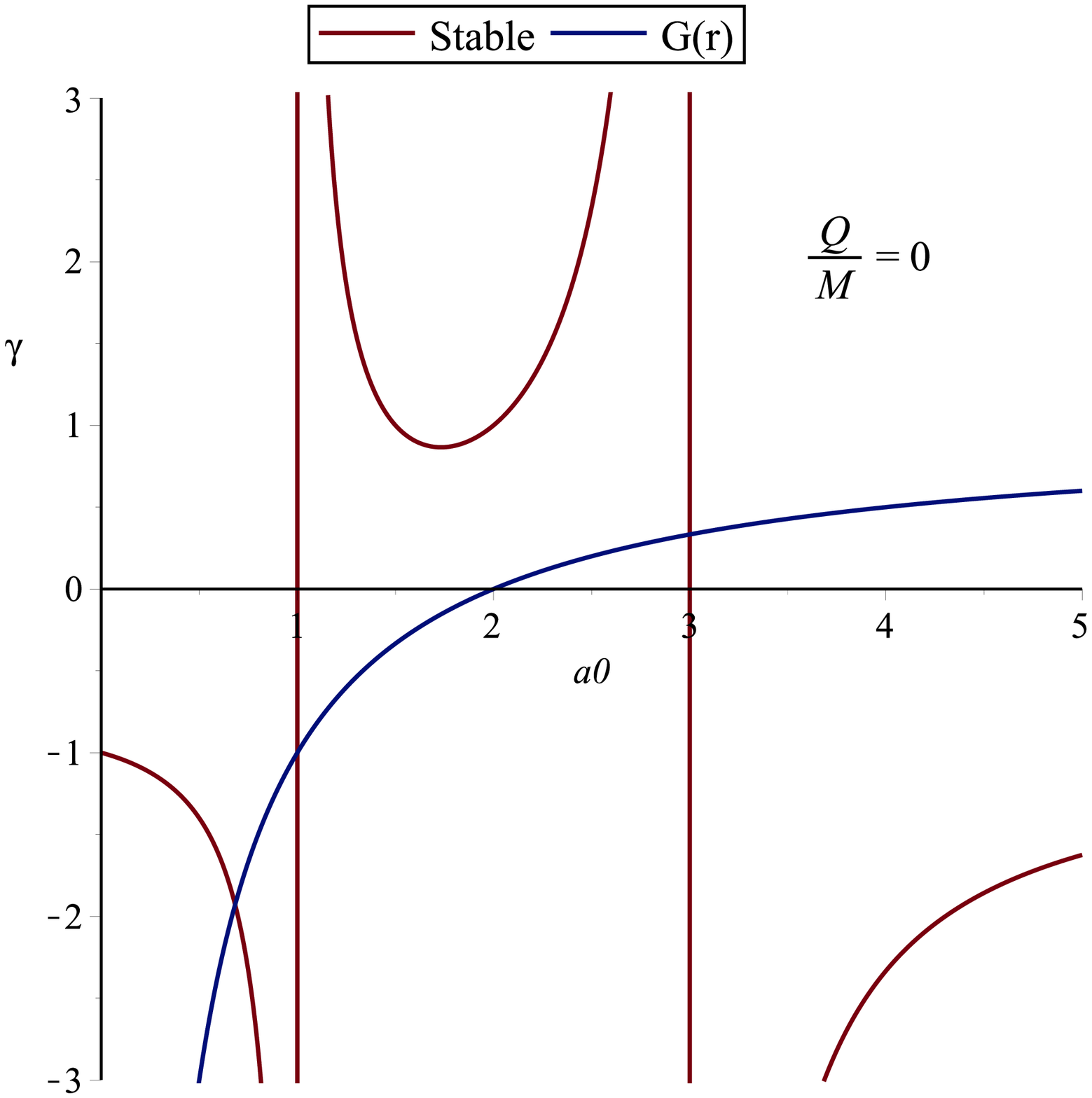,width=0.45\linewidth}\epsfig{file=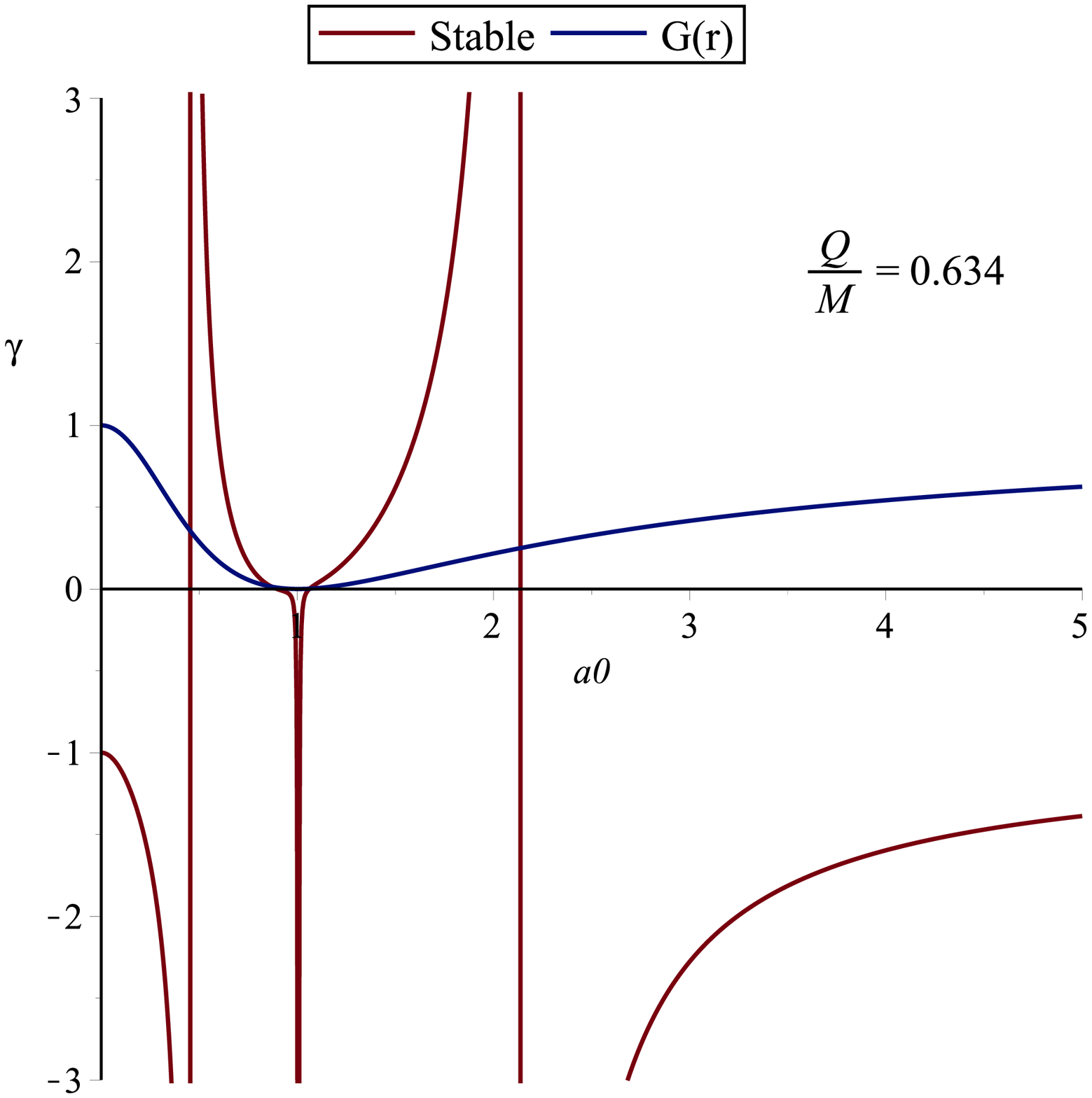,
width=0.45\linewidth}\\
\epsfig{file=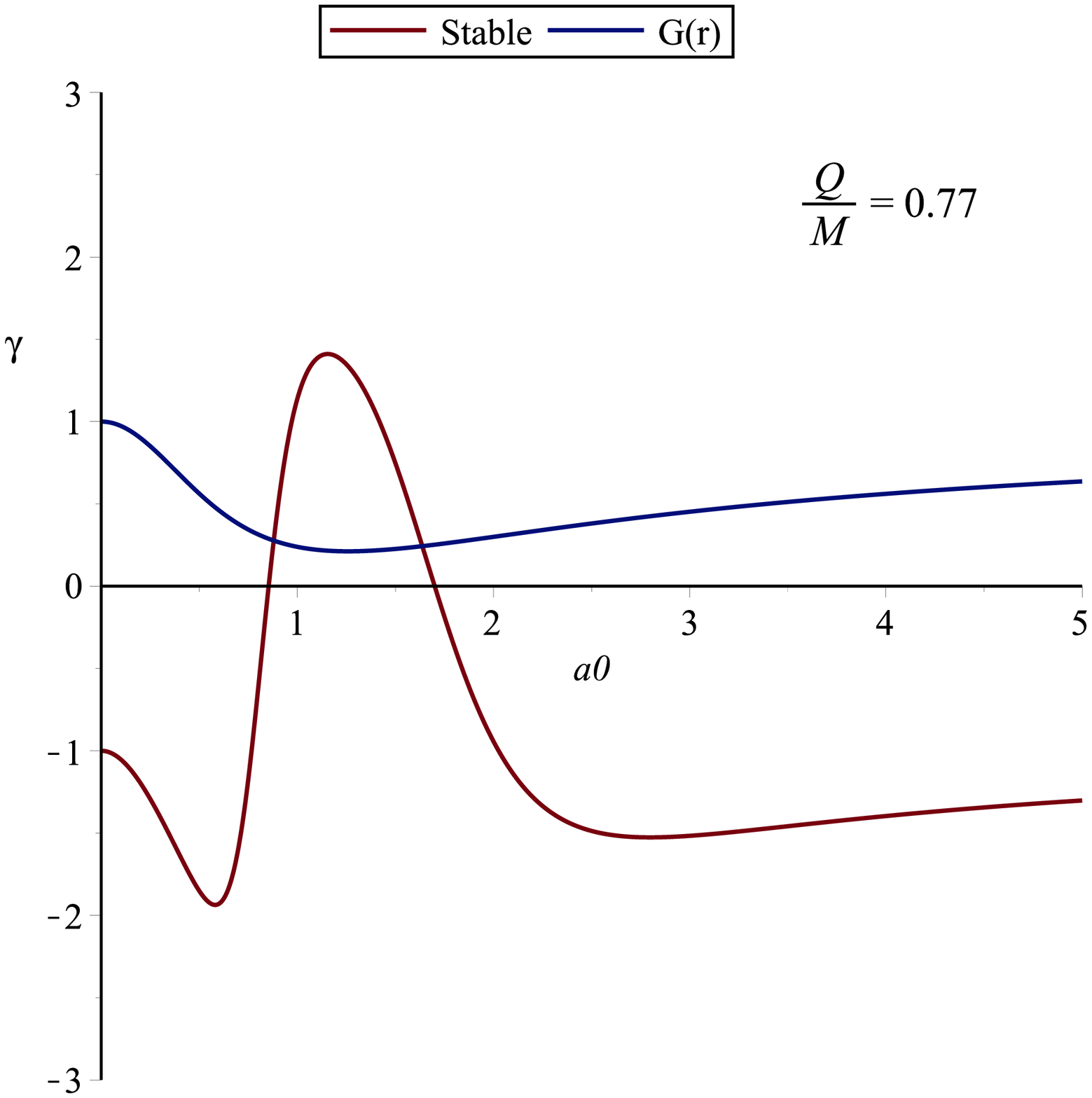,width=0.45\linewidth}\epsfig{file=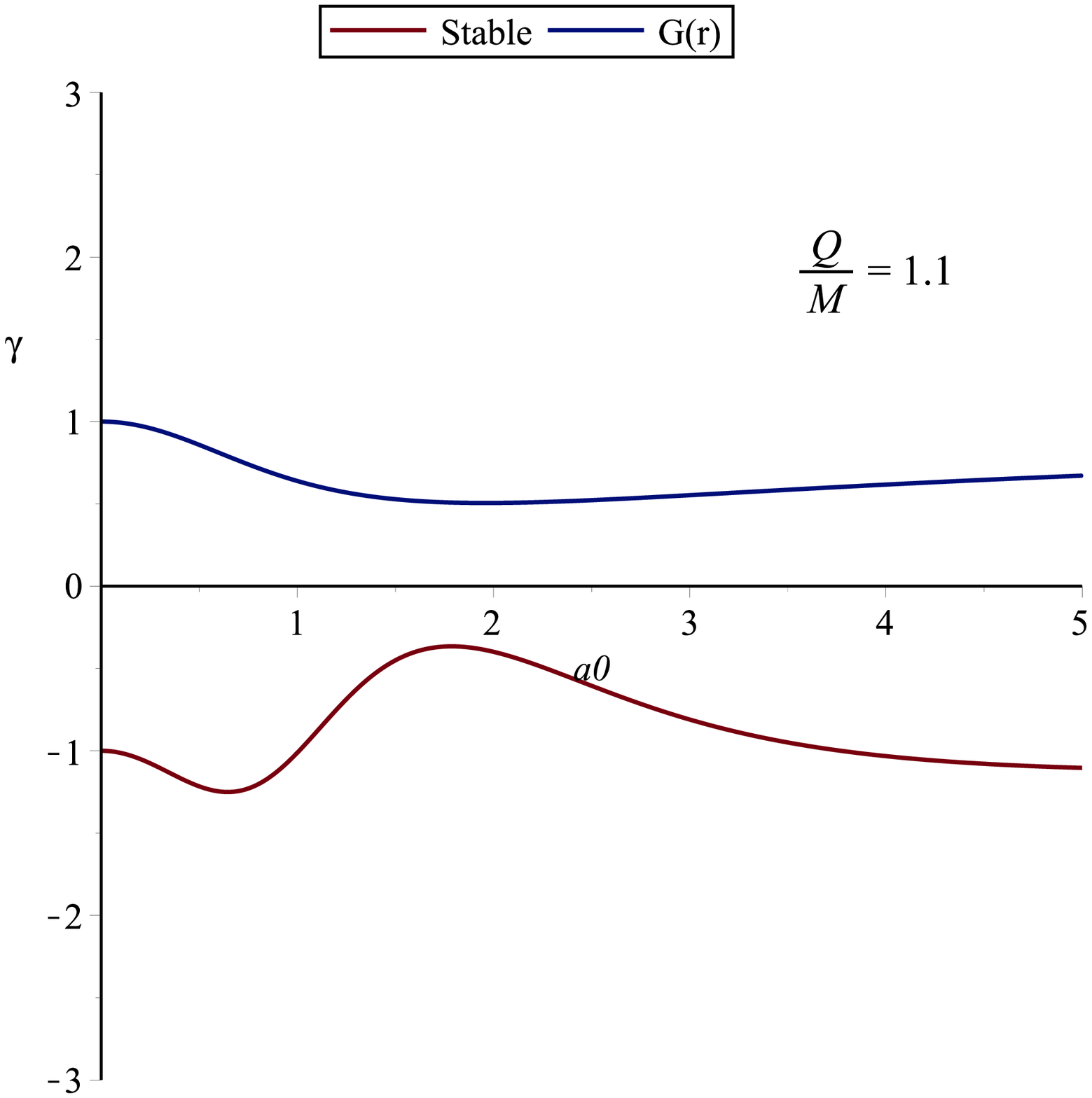,
width=0.45\linewidth}\caption{Plots for stable regular ABG wormholes
with GCG and $\frac{Q}{M}=0, ~0.634, ~0.77, ~1.1$. We plot throat
radius $a_{0}$ and parameter $\gamma$ along abscissa and ordinate,
respectively, where $a_{0}=r$ at thin-shell.}
\end{figure}
\begin{figure}\center
\epsfig{file=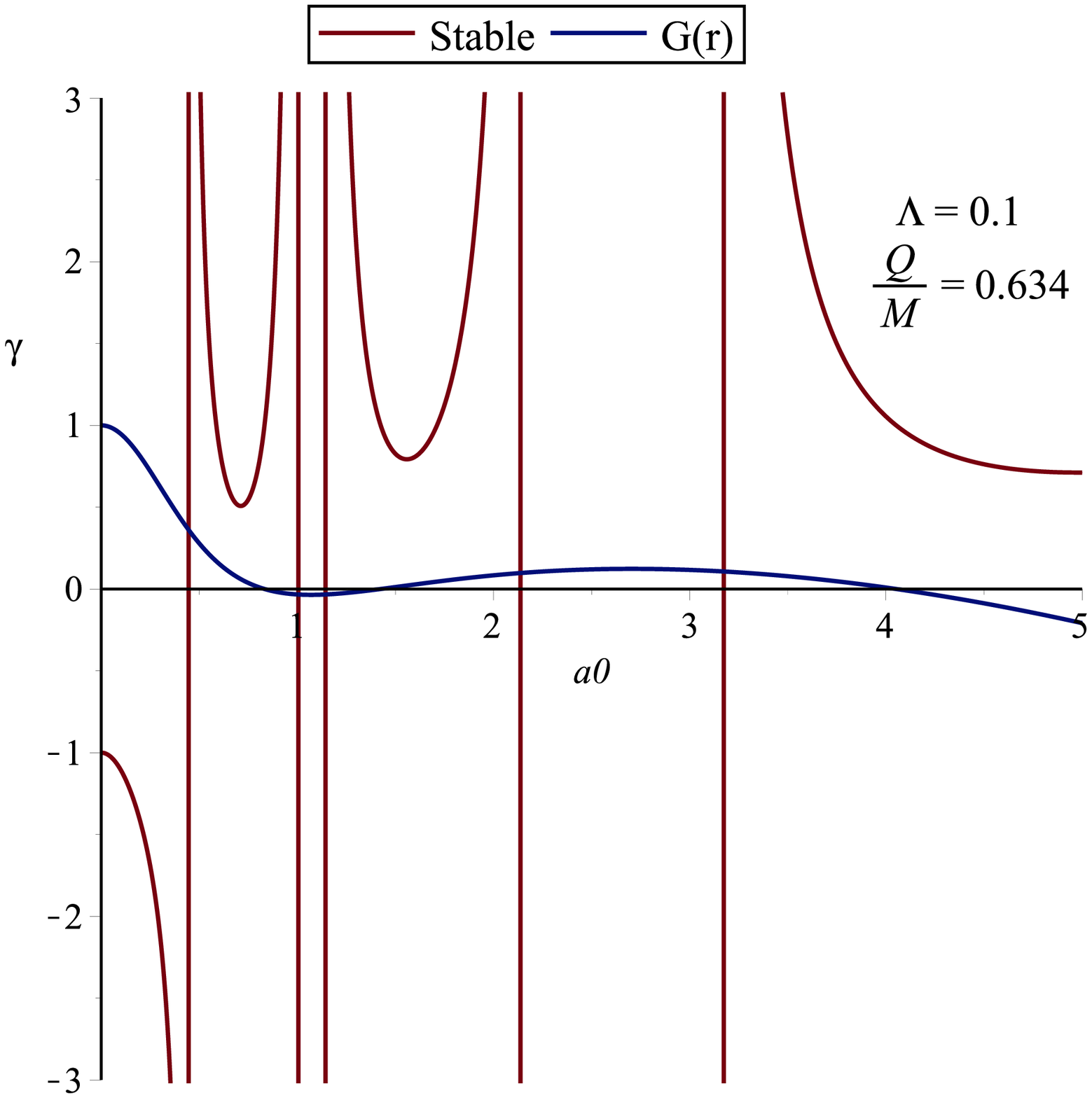,width=0.45\linewidth}\epsfig{file=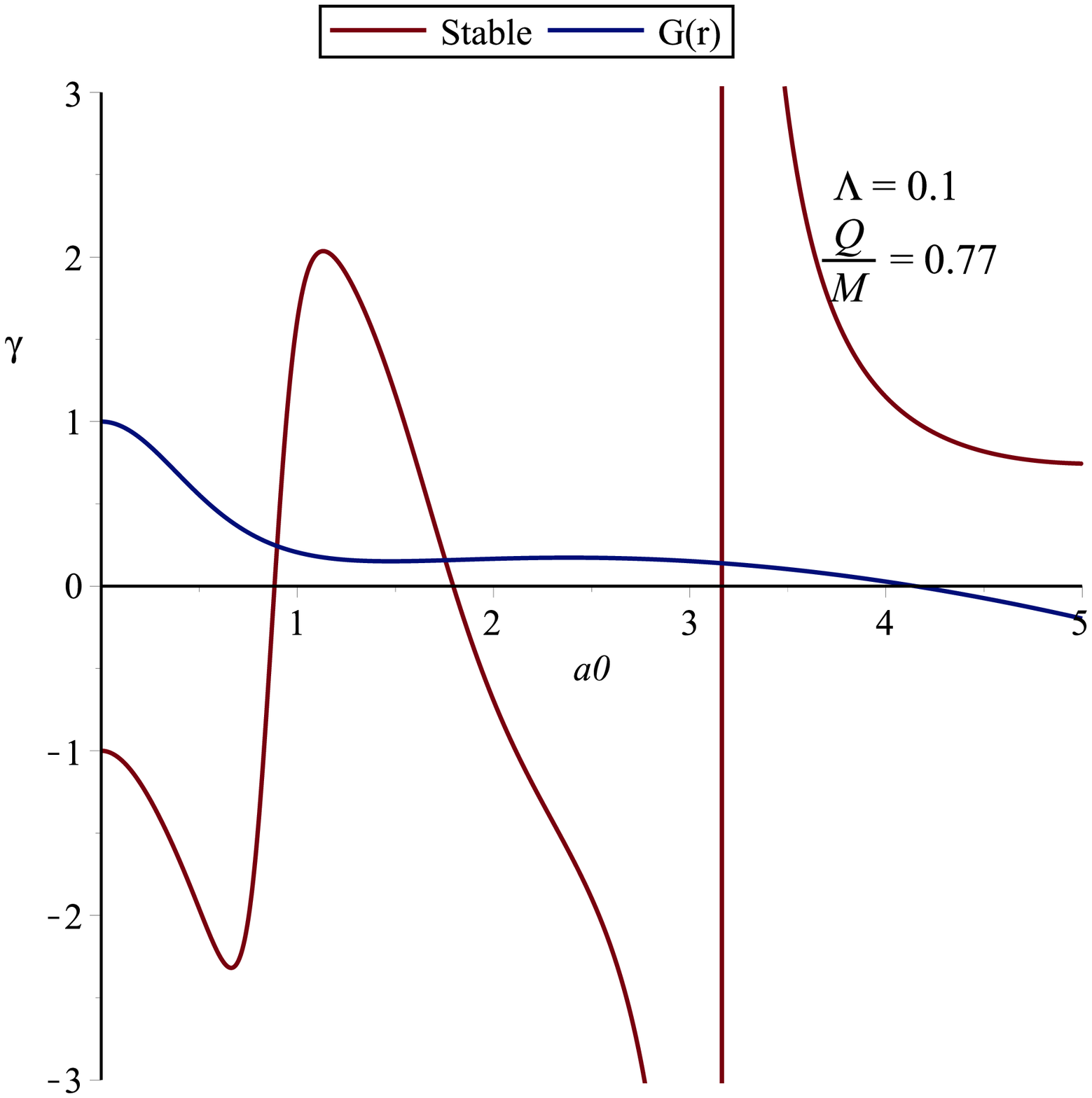,
width=0.45\linewidth}\\
\epsfig{file=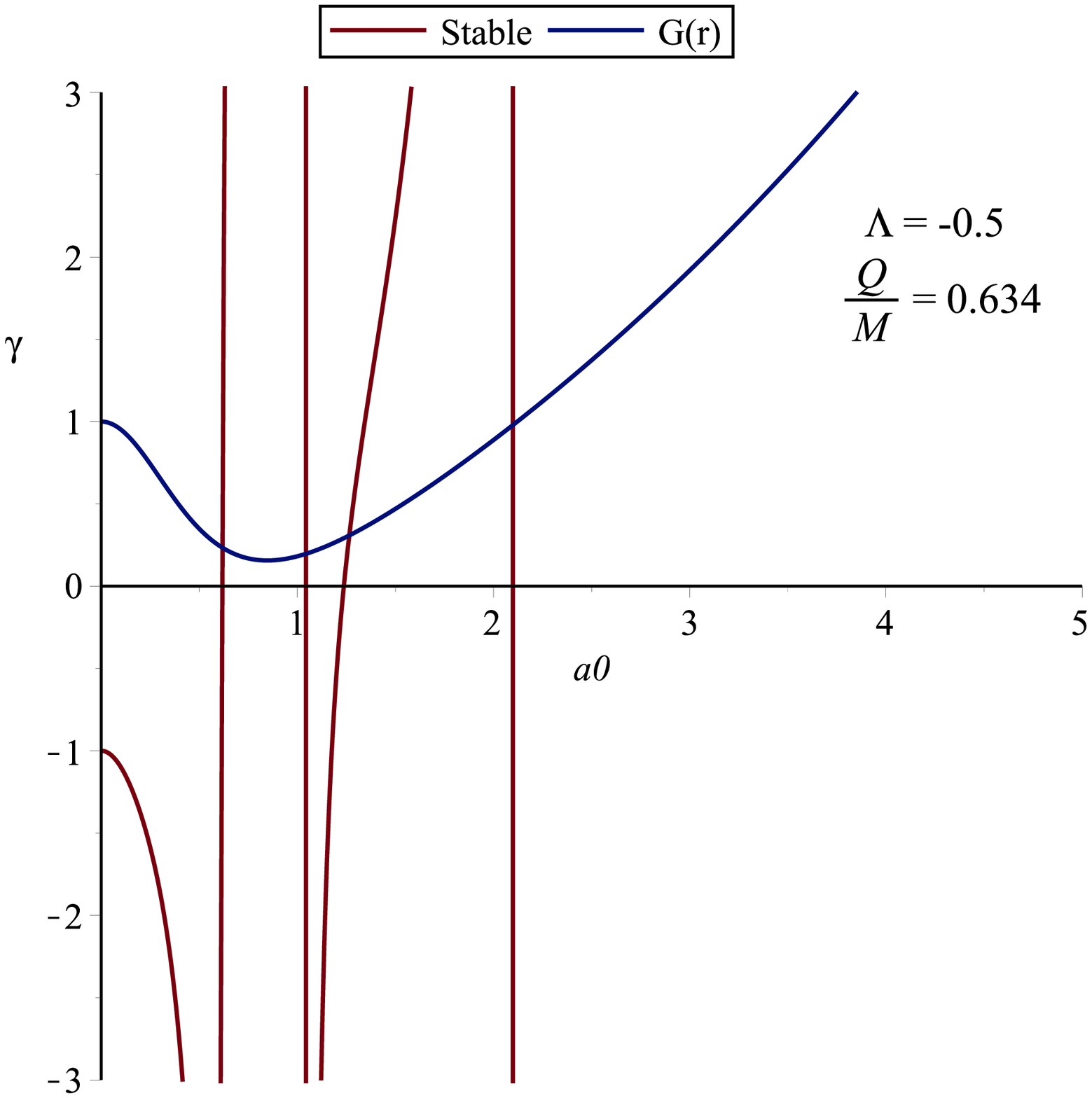,width=0.45\linewidth}\epsfig{file=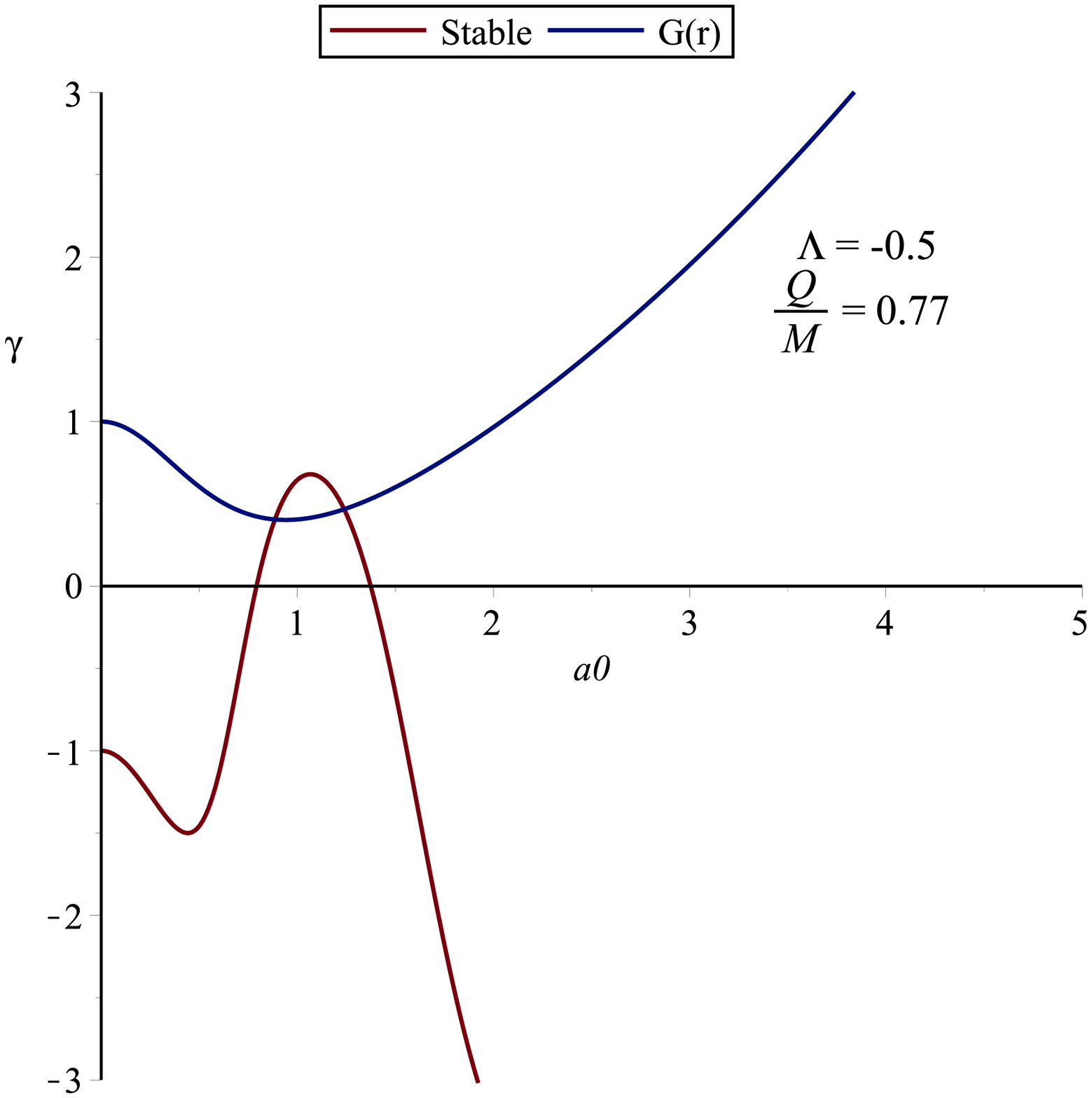,
width=0.45\linewidth}\caption{Plots for stability of regular ABG
wormhole solutions with GCG and $\Lambda=0.1, -0.5$.}
\end{figure}

Here we take GCG model for which EoS is given by
\begin{equation}\label{31}
\Phi(\sigma)=p_{0}+\mu\left(\frac{1}{\sigma^\gamma}-
\frac{1}{\sigma_{0}^\gamma}\right),
\end{equation}
where $\gamma$ is EoS parameter. We are interested to check its
effect on the wormhole stability. For this purpose, we choose
$\mu=p_{0}\sigma^{\gamma}$ which makes the above EoS in the form
\begin{equation}\label{31a}
\Phi(\sigma)=p_{0}\left(\frac{\sigma_{0}}{\sigma}\right)^{\gamma},
\end{equation}
such that $\Phi'(\sigma_{0})=-\frac{p_{0}}{\sigma_{0}}\gamma$. We
plot the respective stability regions numerically as shown in Figure
\textbf{8}. For GCG, we analyze maximum stability areas for
$\frac{Q}{M}=0.634$ which are decreased by increasing $\frac{Q}{M}$
such that only one stable region is obtained for $\frac{Q}{M}=1.1$.
We also plot results for the stability of the respective wormholes
for $\Lambda=0.1,-0.5$ as shown in Figure \textbf{9}. These results
show that more stable regions exist as compared to the case without
$\Lambda$. It is noted that GCG has remarkable importance to provide
maximum stability regions for de Sitter ABG wormholes.

\subsection{Modified Generalized Chaplygin Gas}
\begin{figure}\center
\epsfig{file=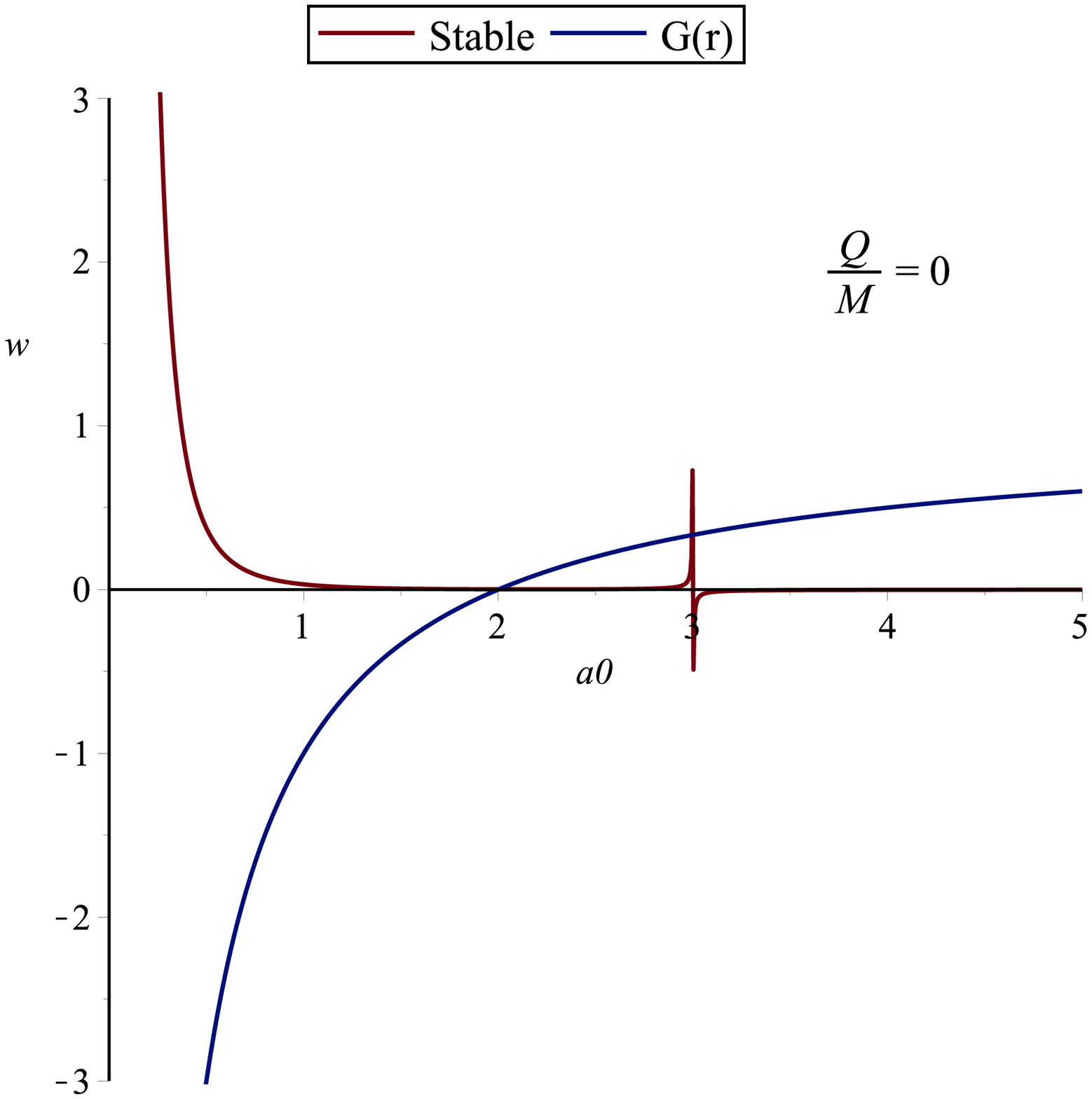,width=0.45\linewidth}\epsfig{file=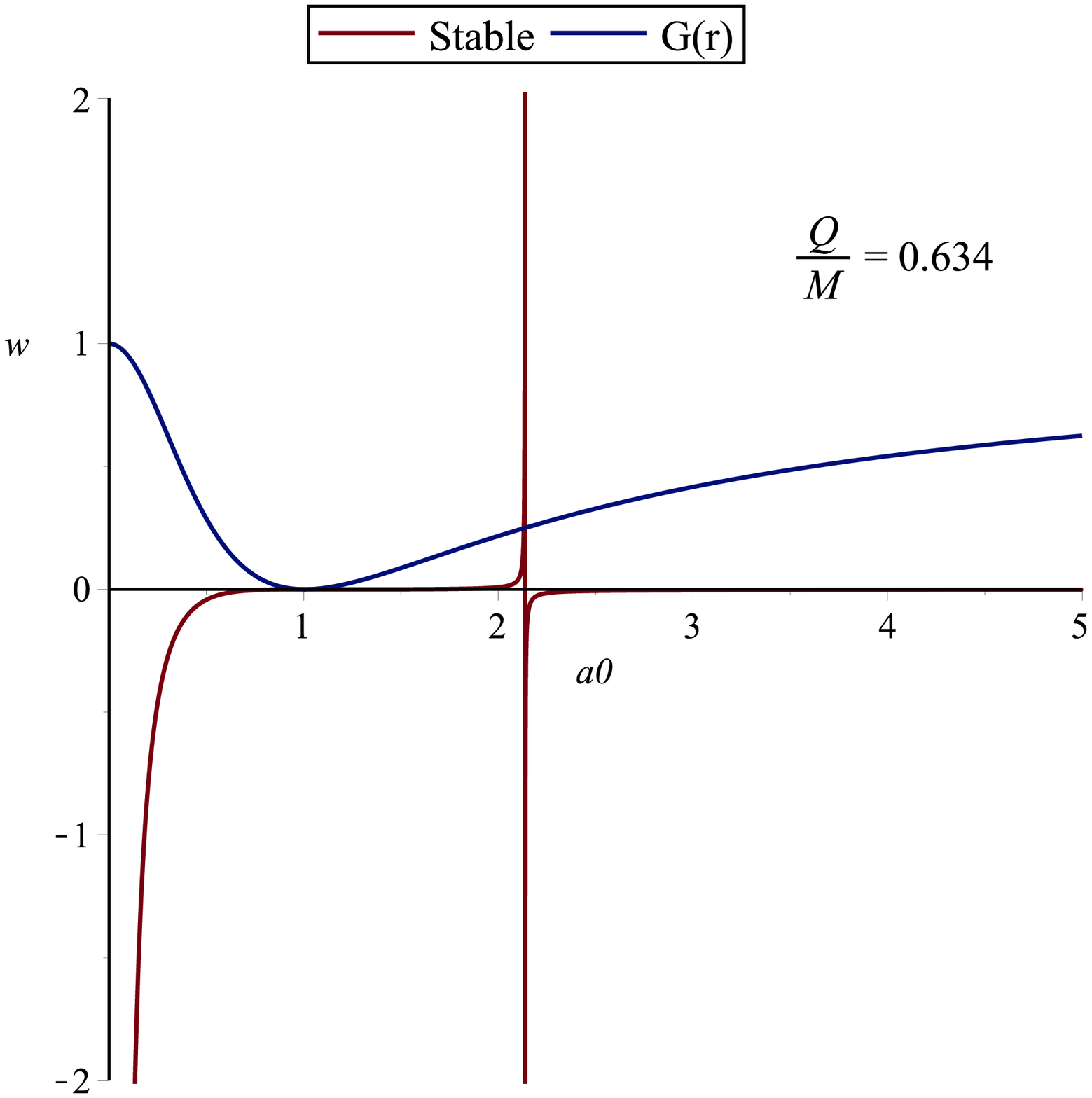,
width=0.45\linewidth}\\
\epsfig{file=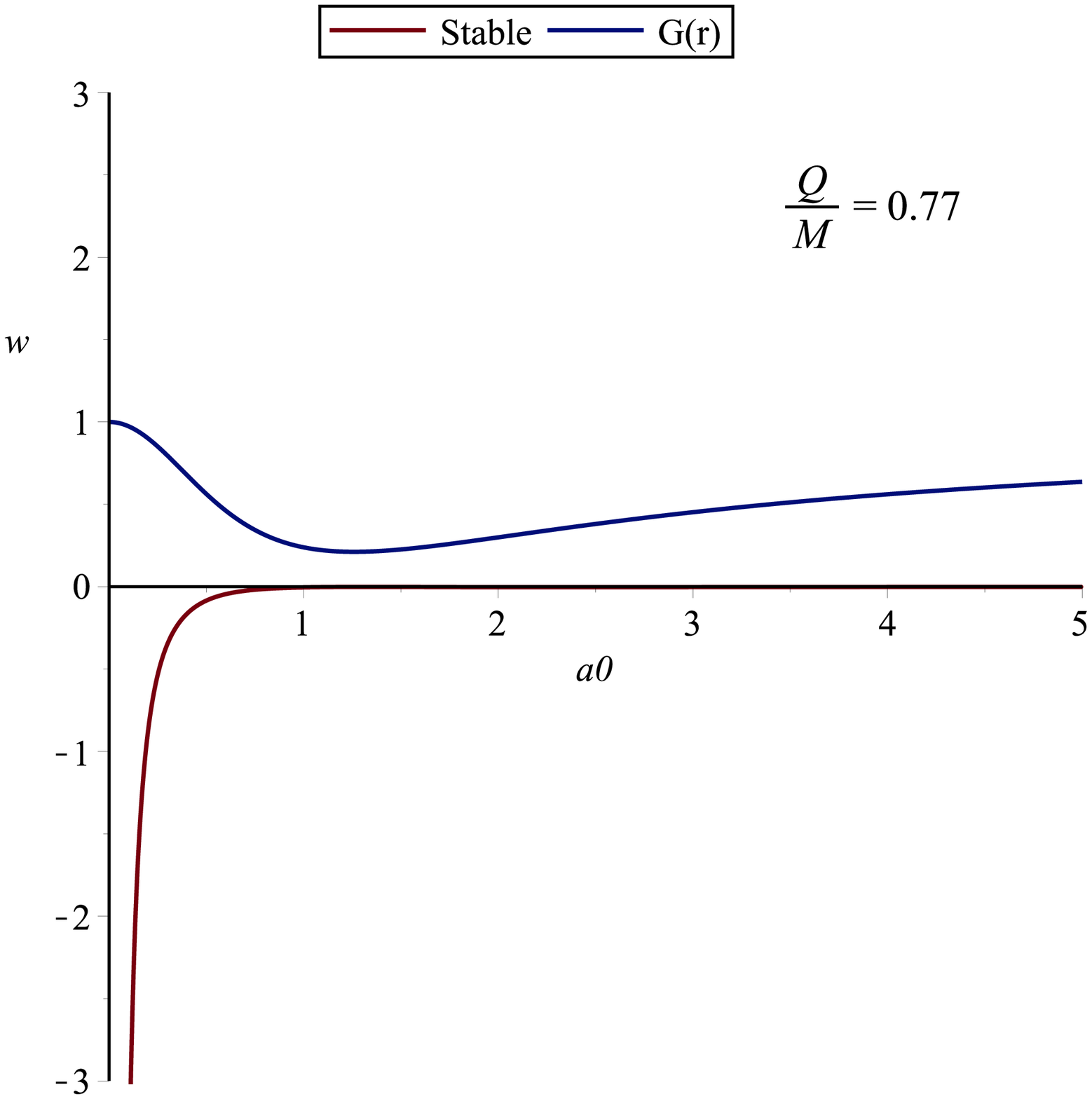,width=0.45\linewidth}\epsfig{file=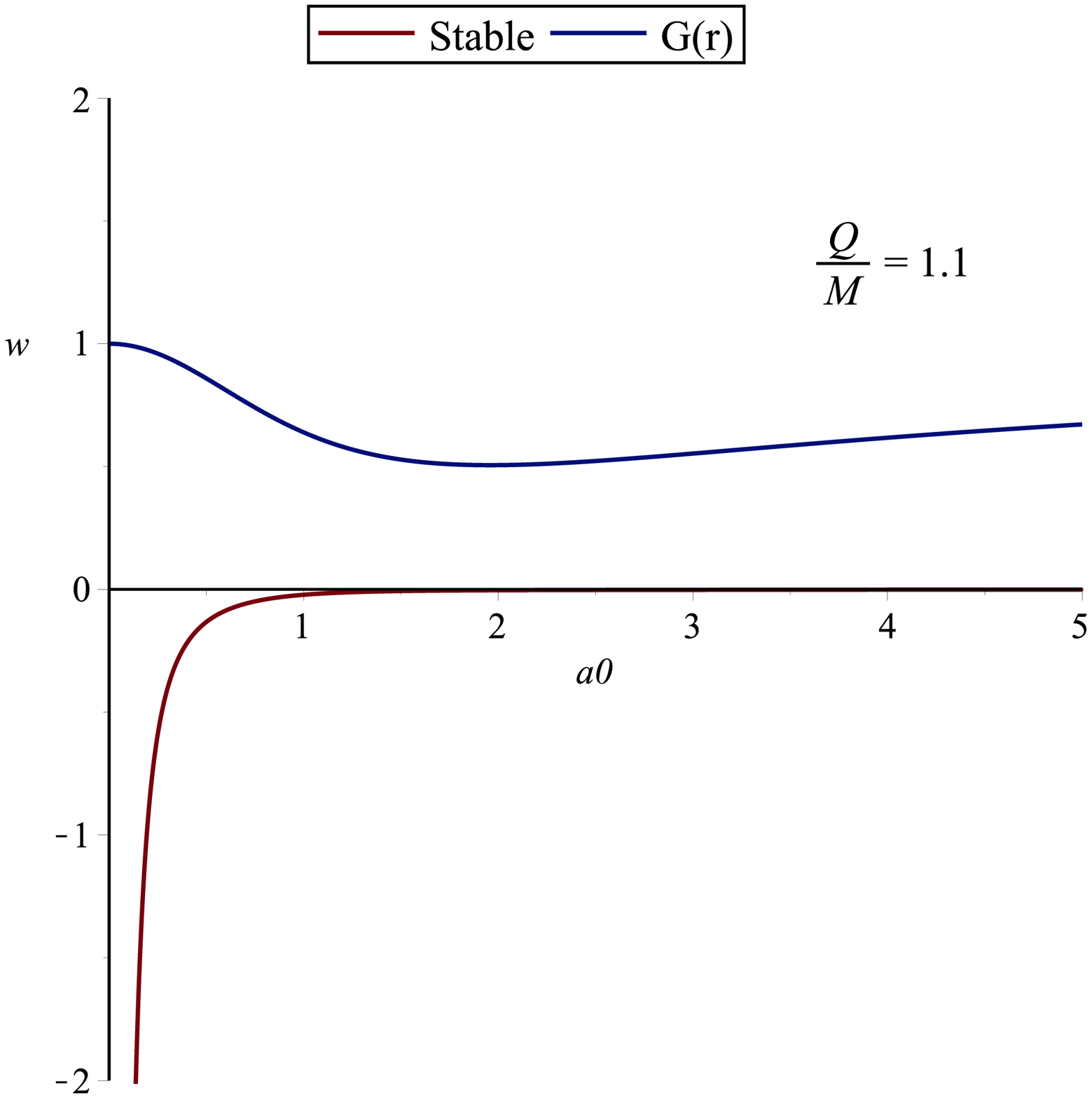,
width=0.45\linewidth}\caption{Plots for stable regular ABG wormholes
in the context of MGCG with $\xi_{0}=\gamma=1$ and different values
of $Q$. We plot throat radius $a_{0}$ and parameter $w$ along
abscissa and ordinate, respectively, where $a_{0}=r$ at thin-shell.}
\end{figure}
\begin{figure}\center
\epsfig{file=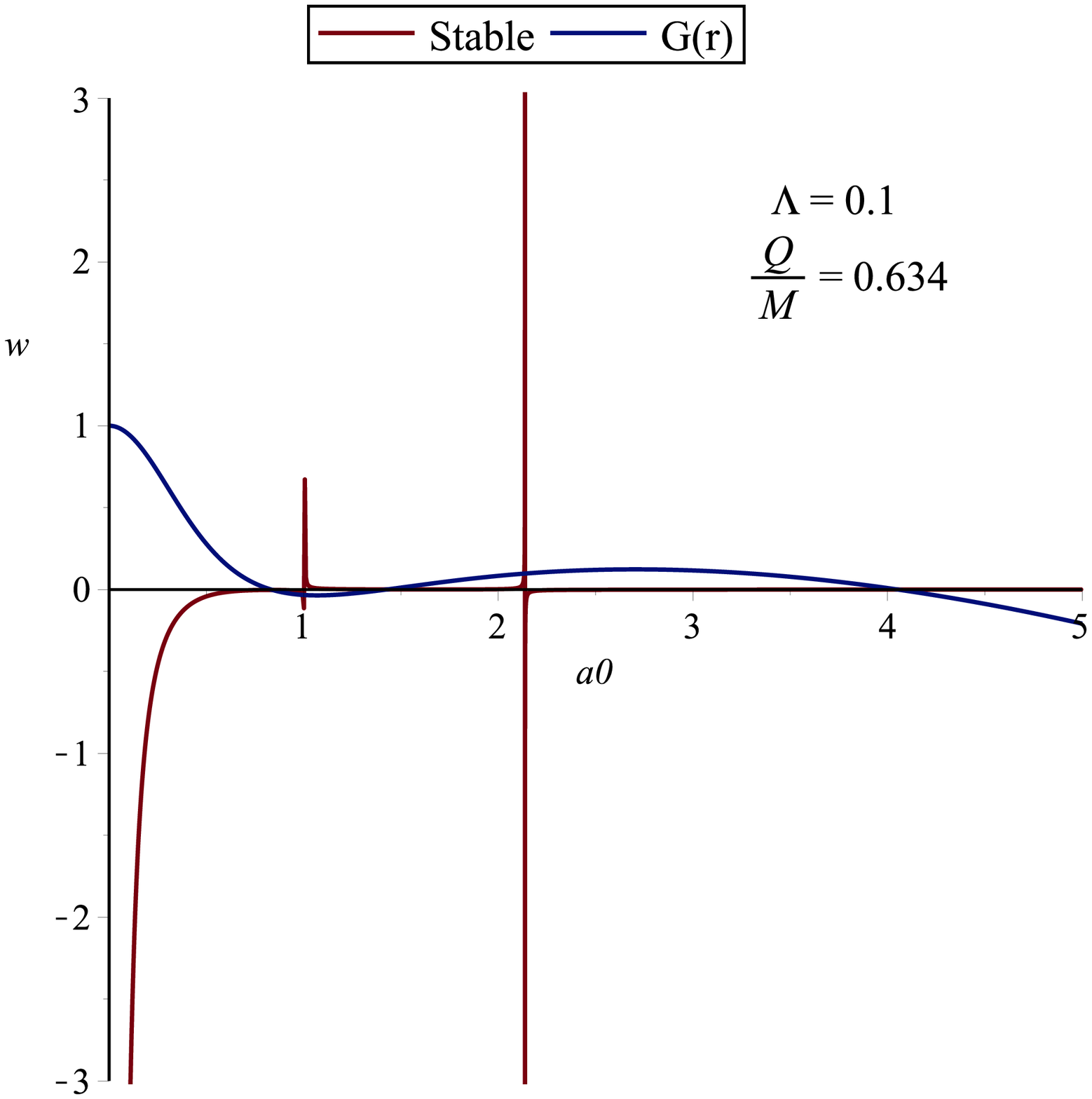,width=0.45\linewidth}\epsfig{file=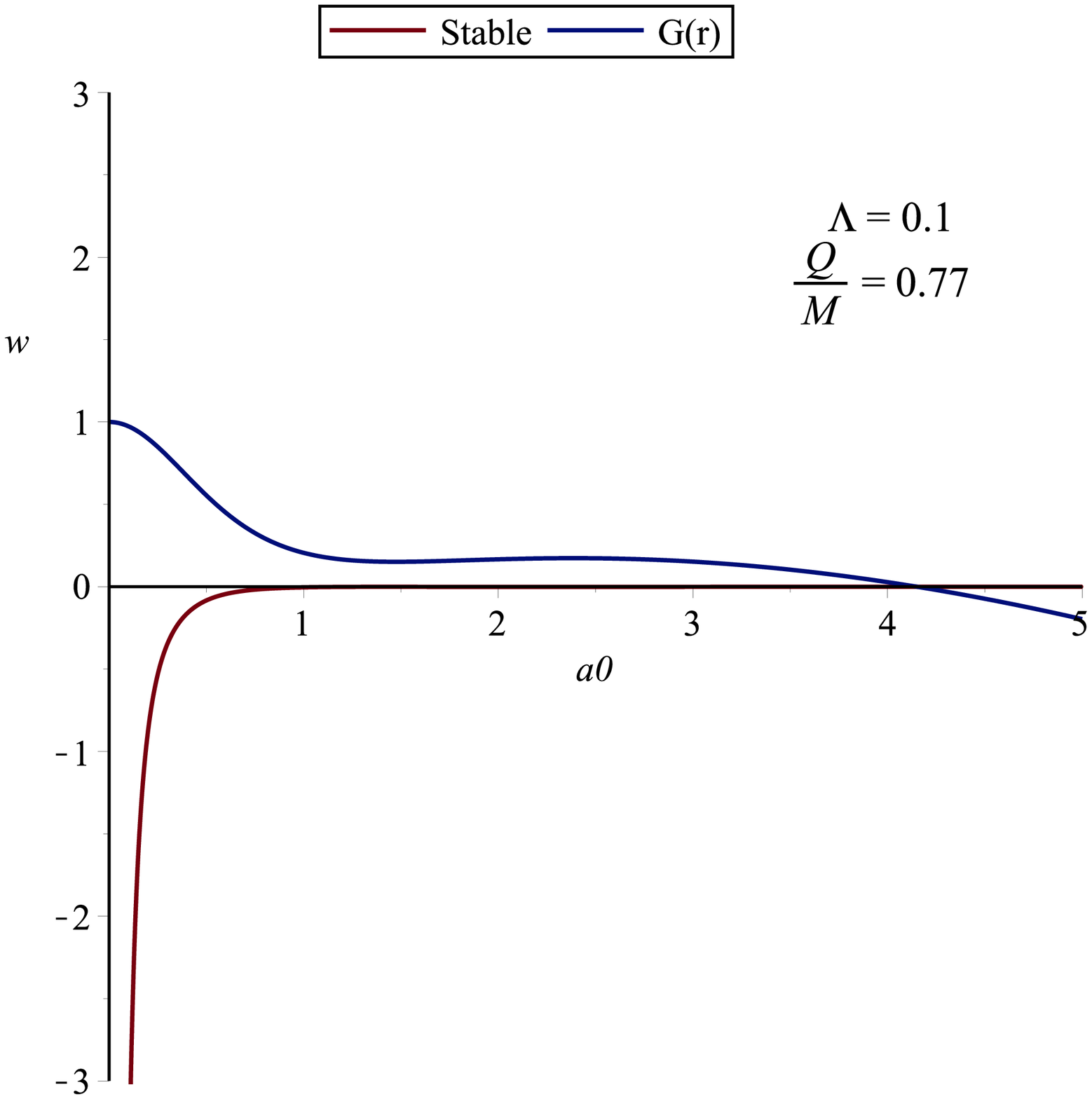,
width=0.45\linewidth}\\
\epsfig{file=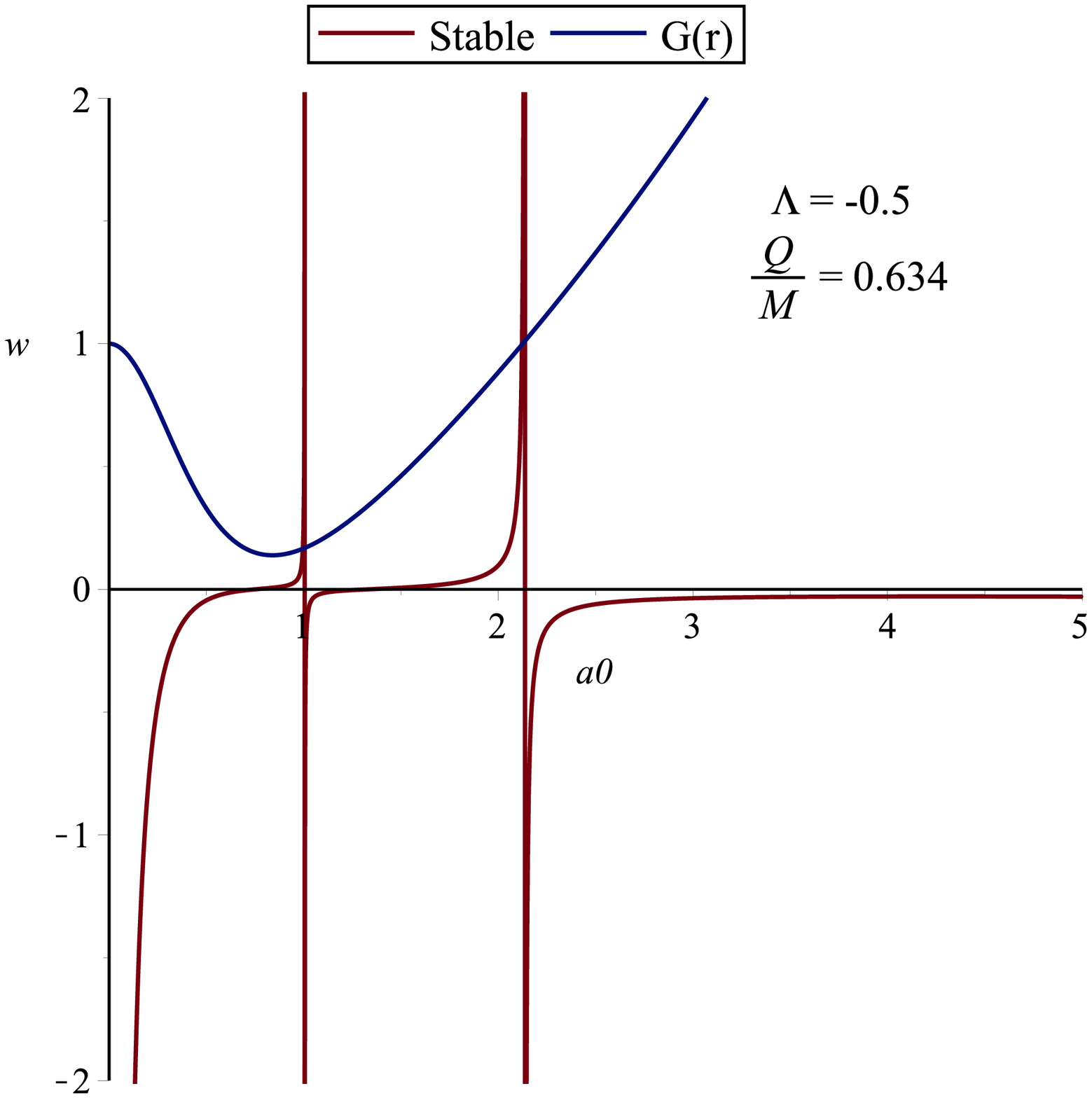,width=0.45\linewidth}\epsfig{file=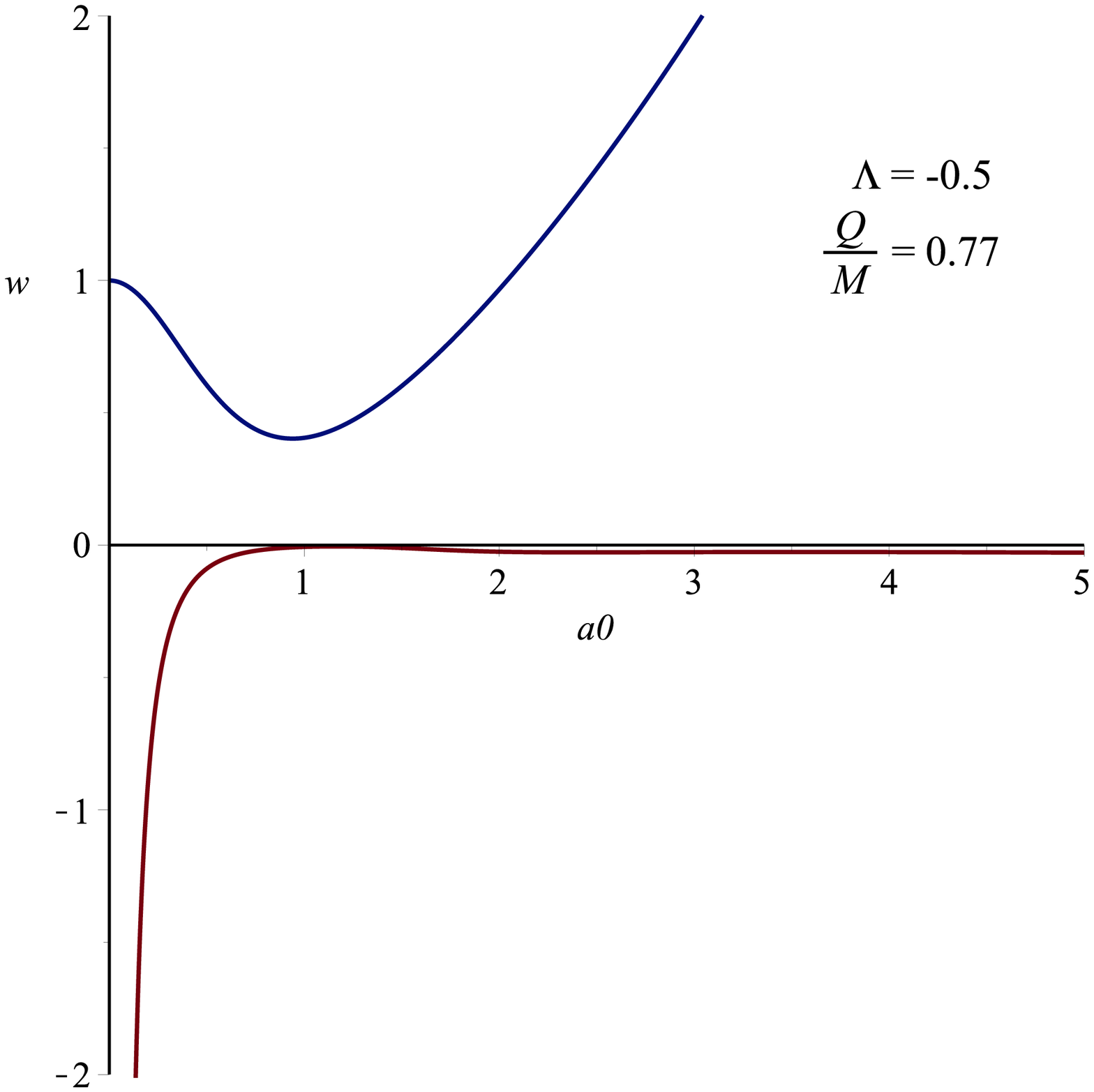,
width=0.45\linewidth}\caption{Plots for stable regular ABG wormhole
solutions with MGCG, $\xi_{0}=\gamma=1$ and $\Lambda=0.1, -0.5$.}
\end{figure}

It is the well-known modification of GCG defined by an EoS
\begin{equation}\label{32}
\Phi(\sigma)=p_{0}+\xi_{0}(\sigma-\sigma_{0})-w\left
(\frac{1}{\sigma^\gamma}- \frac{1}{\sigma_{0}^\gamma}\right),
\end{equation}
where $\xi_{0}$ is a free parameter. Differentiating it w.r.t
$\sigma$, we have
\begin{equation}\nonumber
\Phi'(\sigma_{0})=\xi_{0}+\frac{w\gamma}{\sigma_{0}^{\gamma+1}}.
\end{equation}
We plot the corresponding graphs in Figure \textbf{10} for
$\xi_{0}=\gamma=1$ and different values of charge. For MGCG, we find
the possibility of more stability regions which decreases gradually
by increasing $\frac{Q}{M}$ and reduces to one stable region for
$\frac{Q}{M}=1.1$. For $\Lambda=0.1,-0.5$, we find that the
increasing value of $\frac{Q}{M}$ decreases the possibility of
stable solutions (Figure \textbf{11}). Here we analyze maximum
stable regions for $\Lambda=-0.5$ and $\frac{Q}{M}=0.634$ which
supports the fact that more stable regions exist in anti-de Sitter
case as compared to de Sitter case.

\subsection{Logarithmic Gas}

Now we consider logarithmic gas governed by the following EoS
\begin{equation}\label{33}
\Phi(\sigma)=p_{0}+w\ln\left|\frac{\sigma} {\sigma_{0}}\right|,
\end{equation}
where $\Phi'(\sigma_{0})=\frac{w}{\sigma_{0}}$. The corresponding
stable ABG thin-shell wormholes for $\frac{Q}{M}=0, ~0.634, ~0.77,
~1.1$ without $\Lambda$ are shown in Figure \textbf{12}. It is found
that more stable wormhole solutions exist for $\frac{Q}{M}=0.634$
and stability area decreases by increasing $\frac{Q}{M}$. We find
only one stable region for larger values of $\frac{Q}{M}$. It is
observed that more stable regions exist for de Sitter and anti-de
Sitter spacetimes in comparison to the spacetime without $\Lambda$
(Figure \textbf{13}). We obtain maximum stable regions for
$\frac{Q}{M}=0.634$ in de Sitter and anti-de Sitter spacetimes. We
investigate that the effect of logarithmic gas with inclusion of
$\Lambda$ is to enhance the stability regions for regular ABG
wormhole configurations as depicted in our numerical plots.
\begin{figure}\center
\epsfig{file=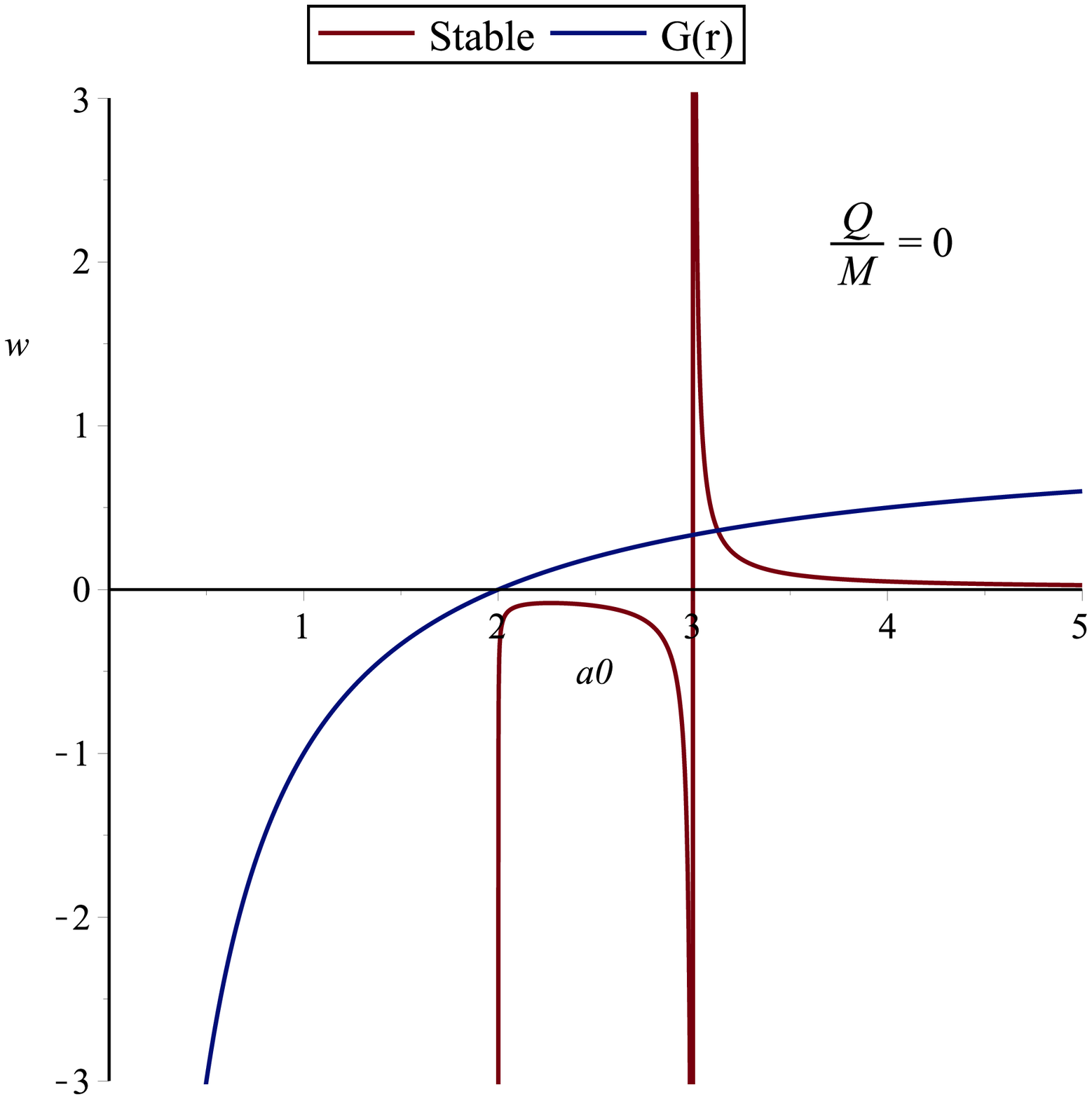,width=0.45\linewidth}\epsfig{file=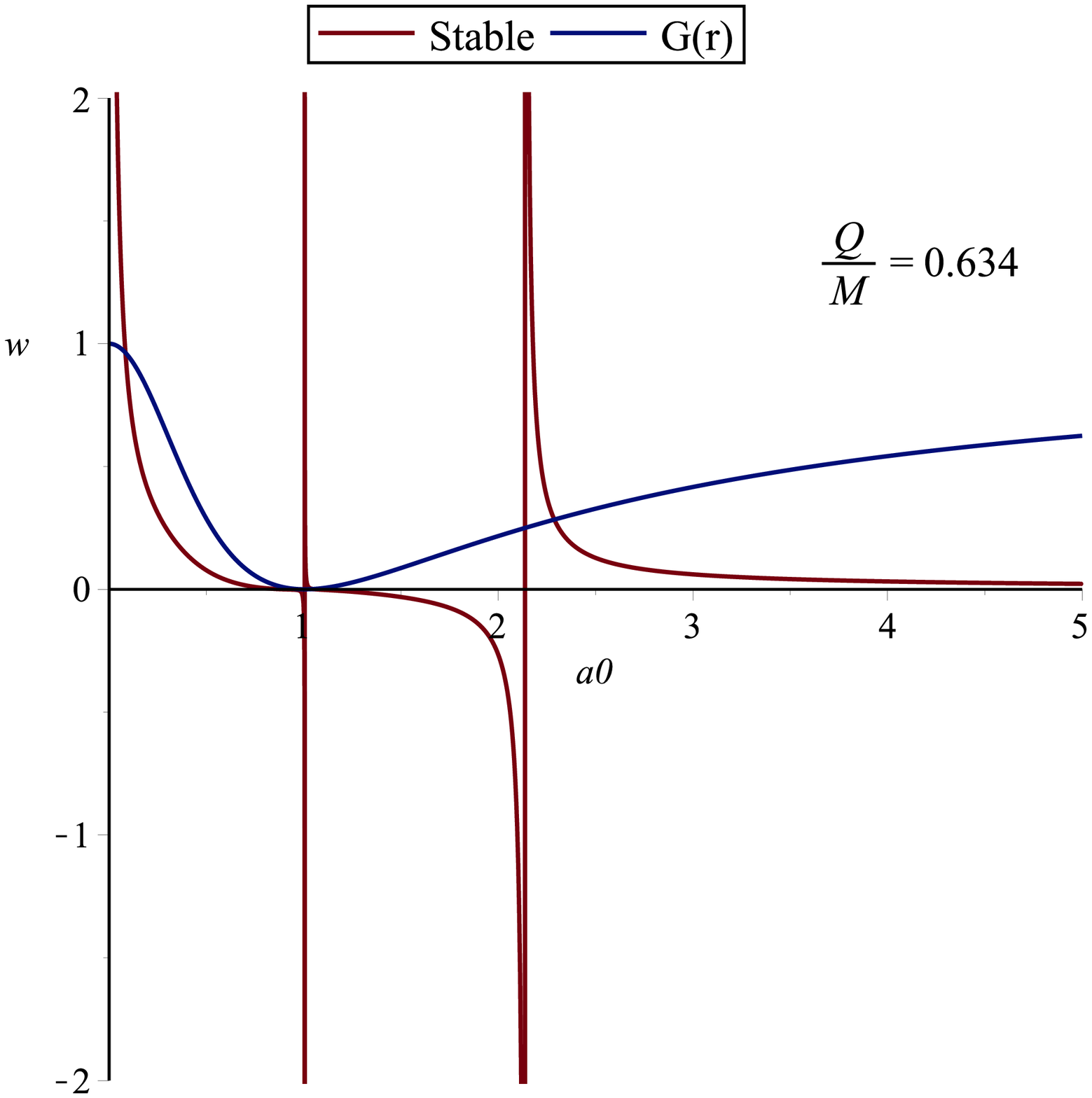,
width=0.45\linewidth}\\
\epsfig{file=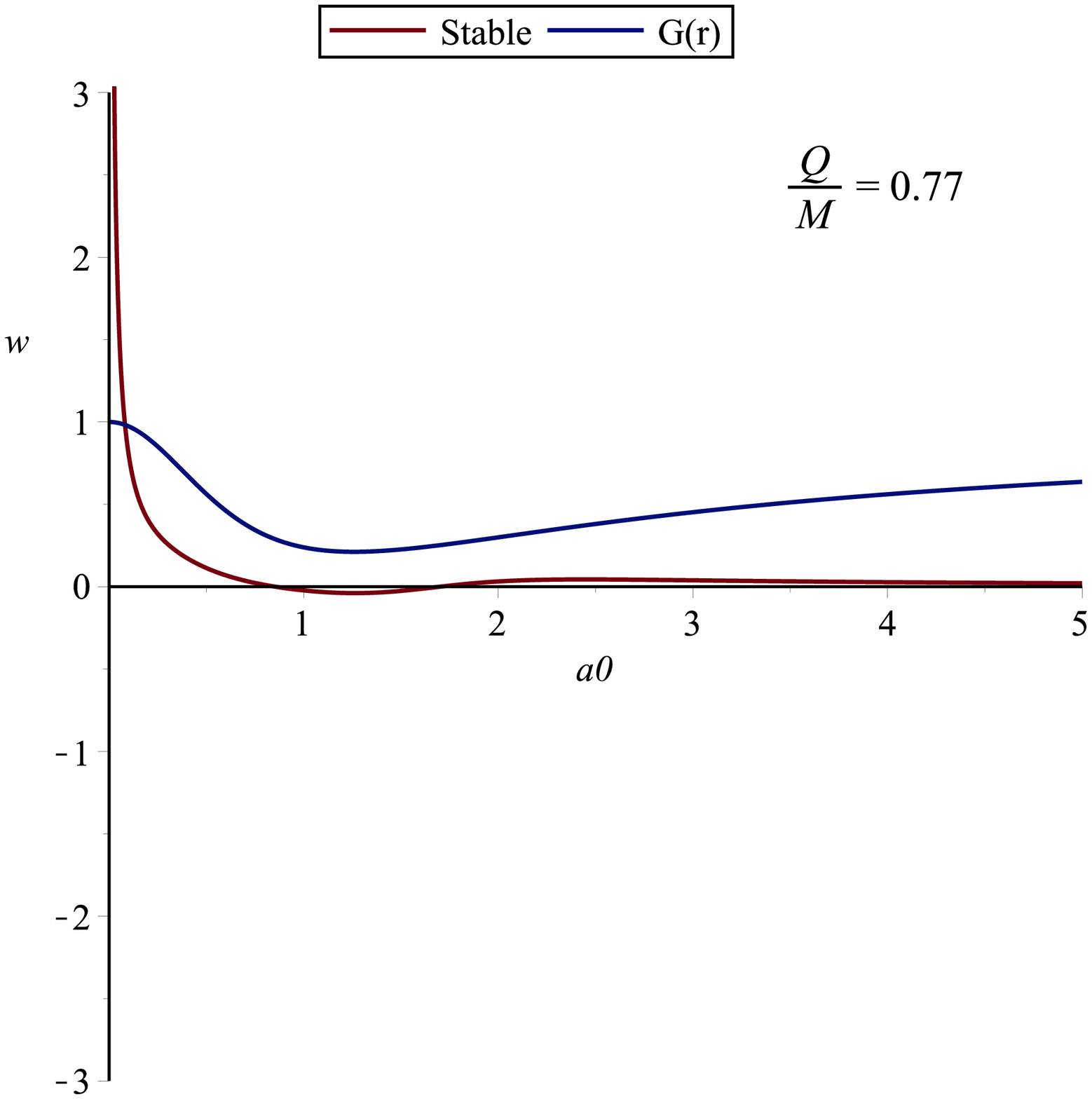,width=0.45\linewidth}\epsfig{file=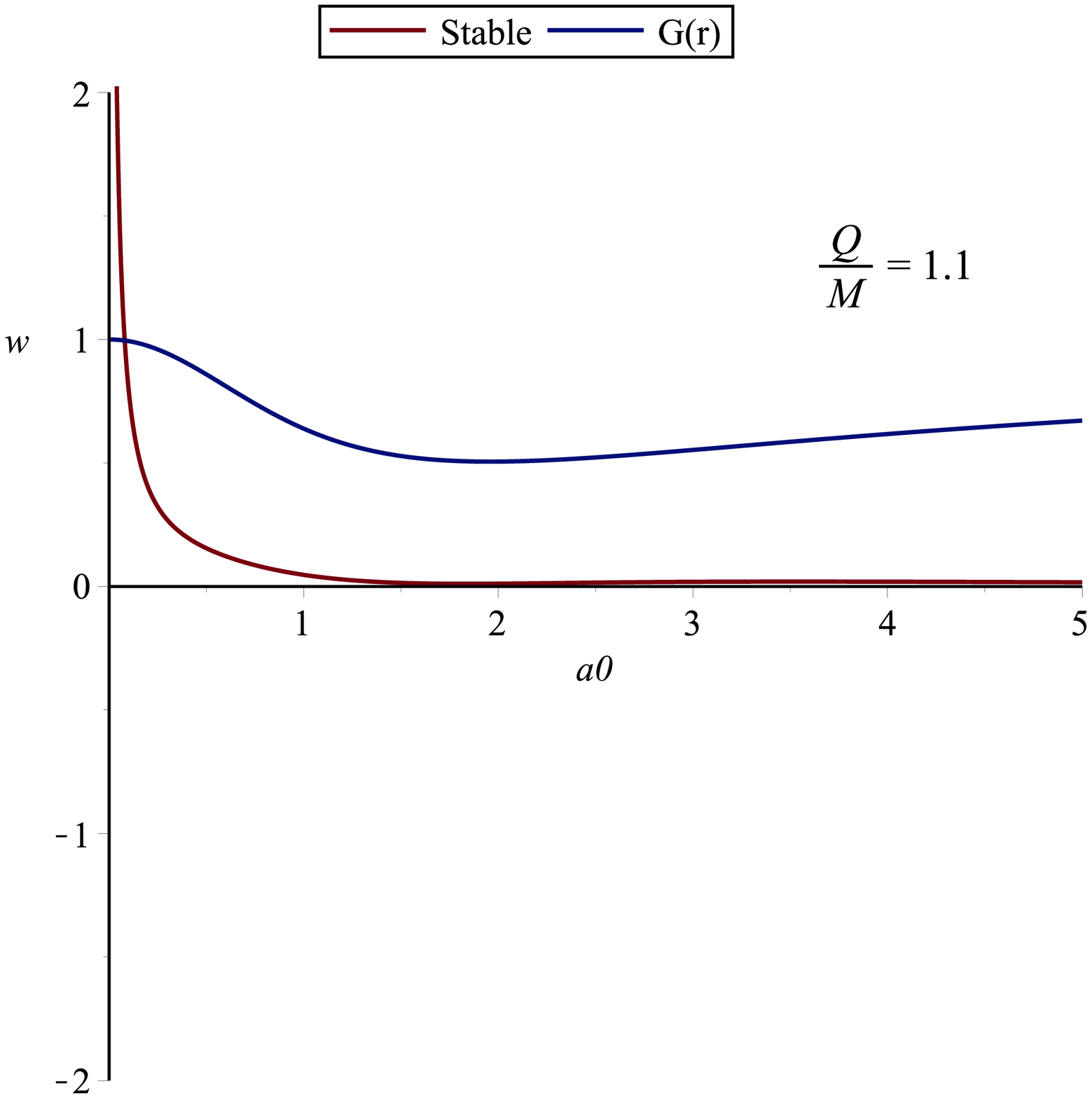,
width=0.45\linewidth}\caption{Plots for logarithmic gas EoS with
$\frac{Q}{M}=0, ~0.634,~0.77, ~1.1$. We plot throat radius $a_{0}$
and parameter $w$ along abscissa and ordinate, respectively, where
$a_{0}=r$ at thin-shell.}
\end{figure}
\begin{figure}\center
\epsfig{file=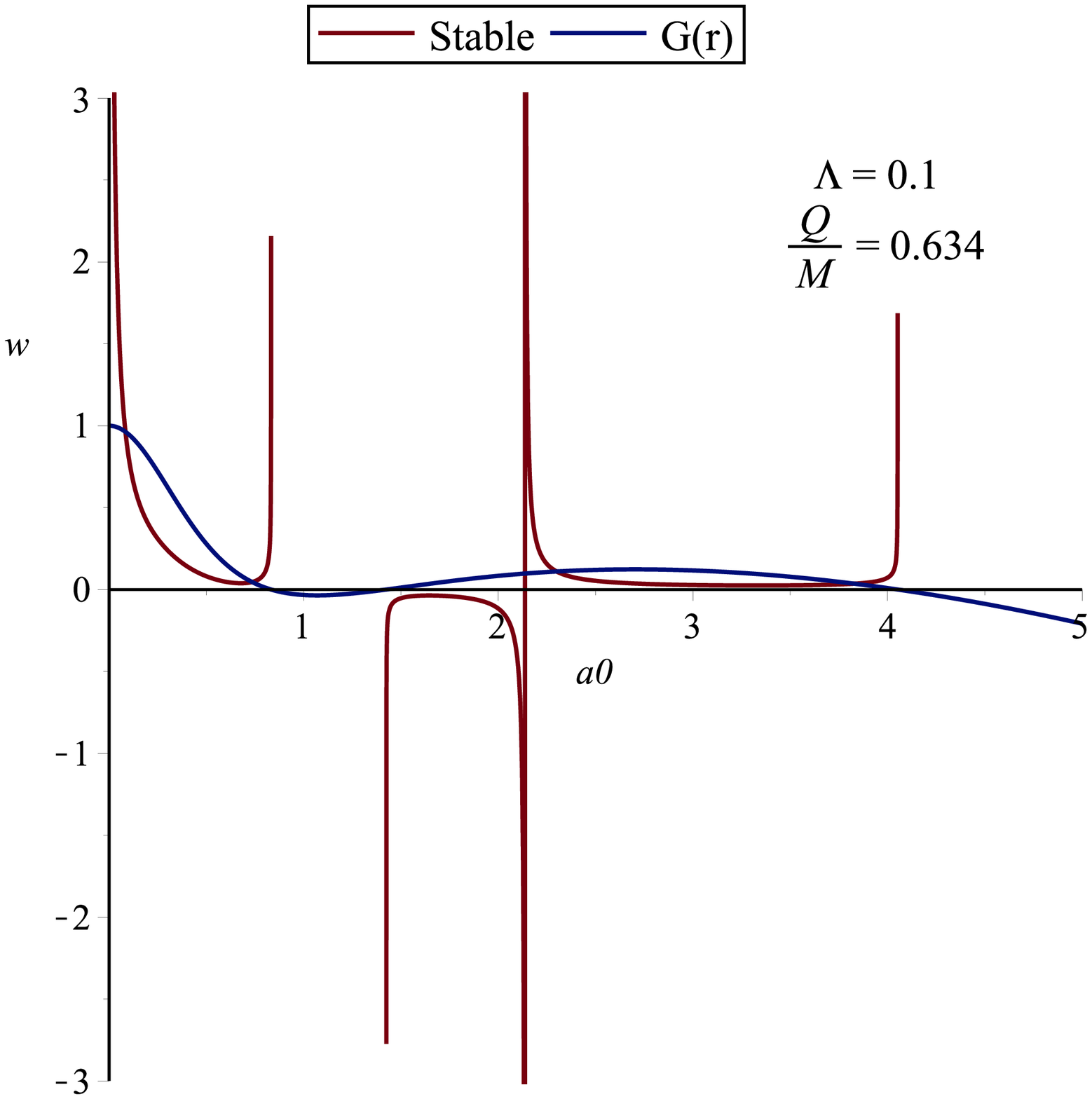,width=0.45\linewidth}\epsfig{file=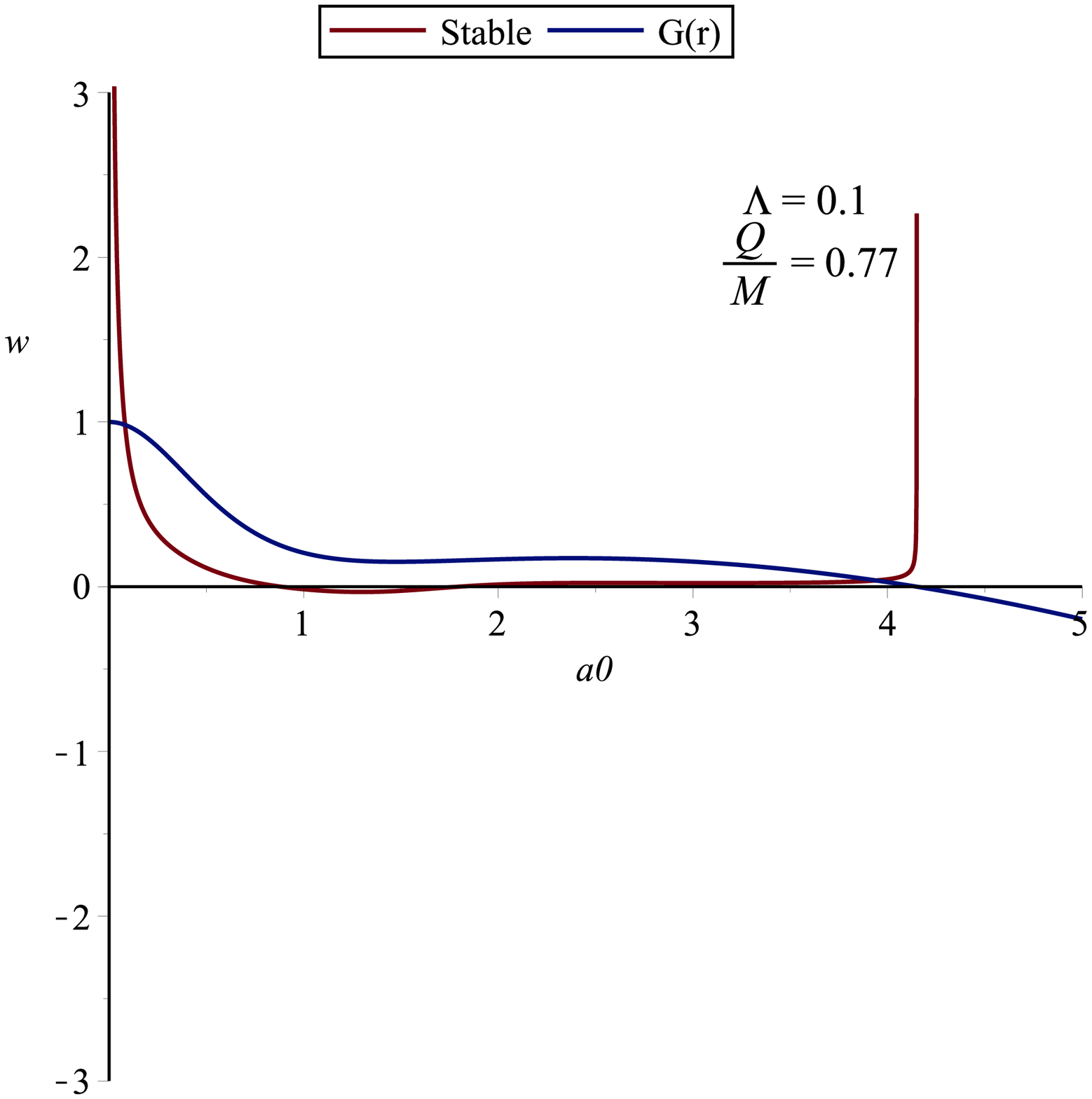,
width=0.45\linewidth}\\
\epsfig{file=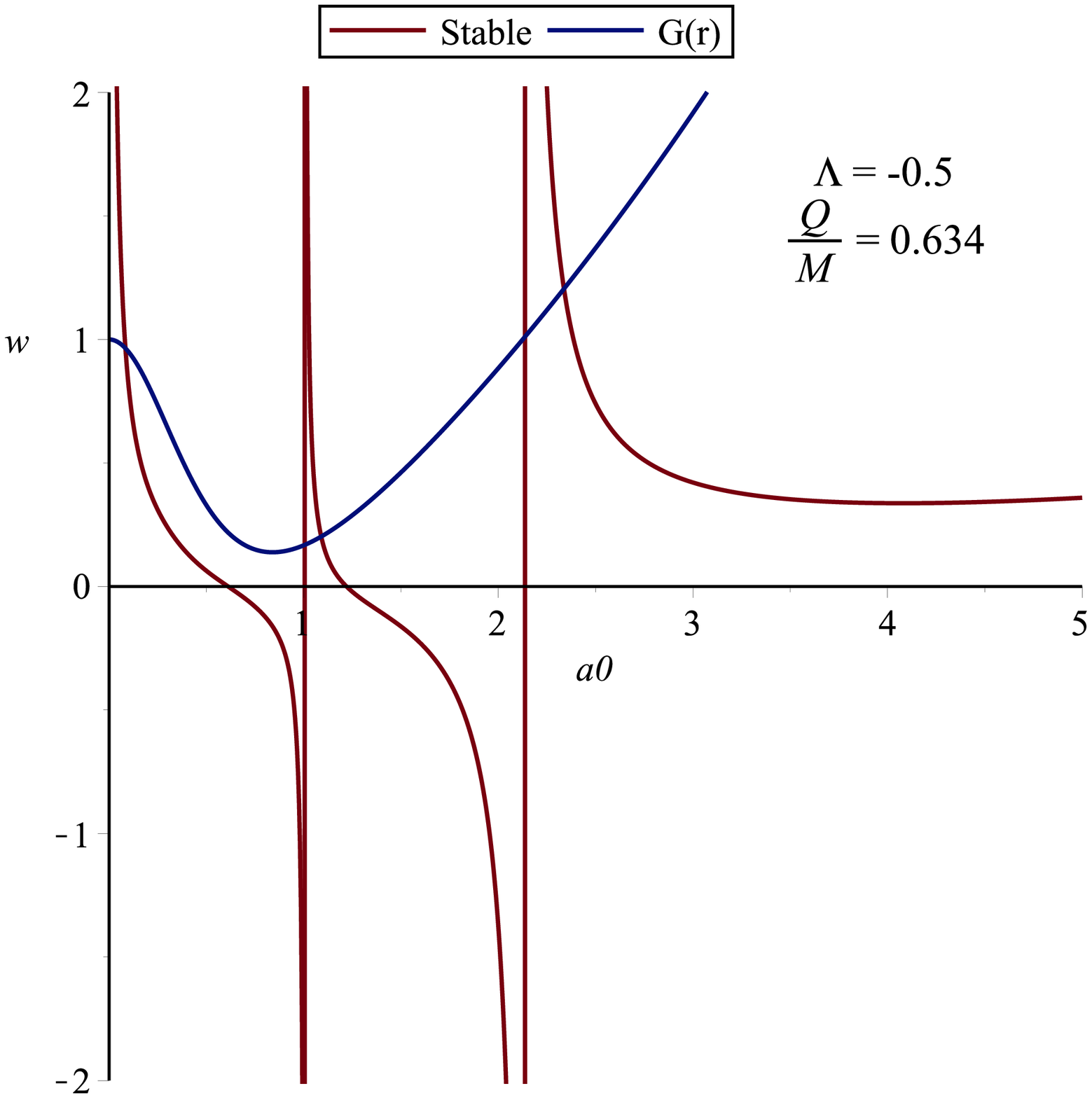,width=0.45\linewidth}\epsfig{file=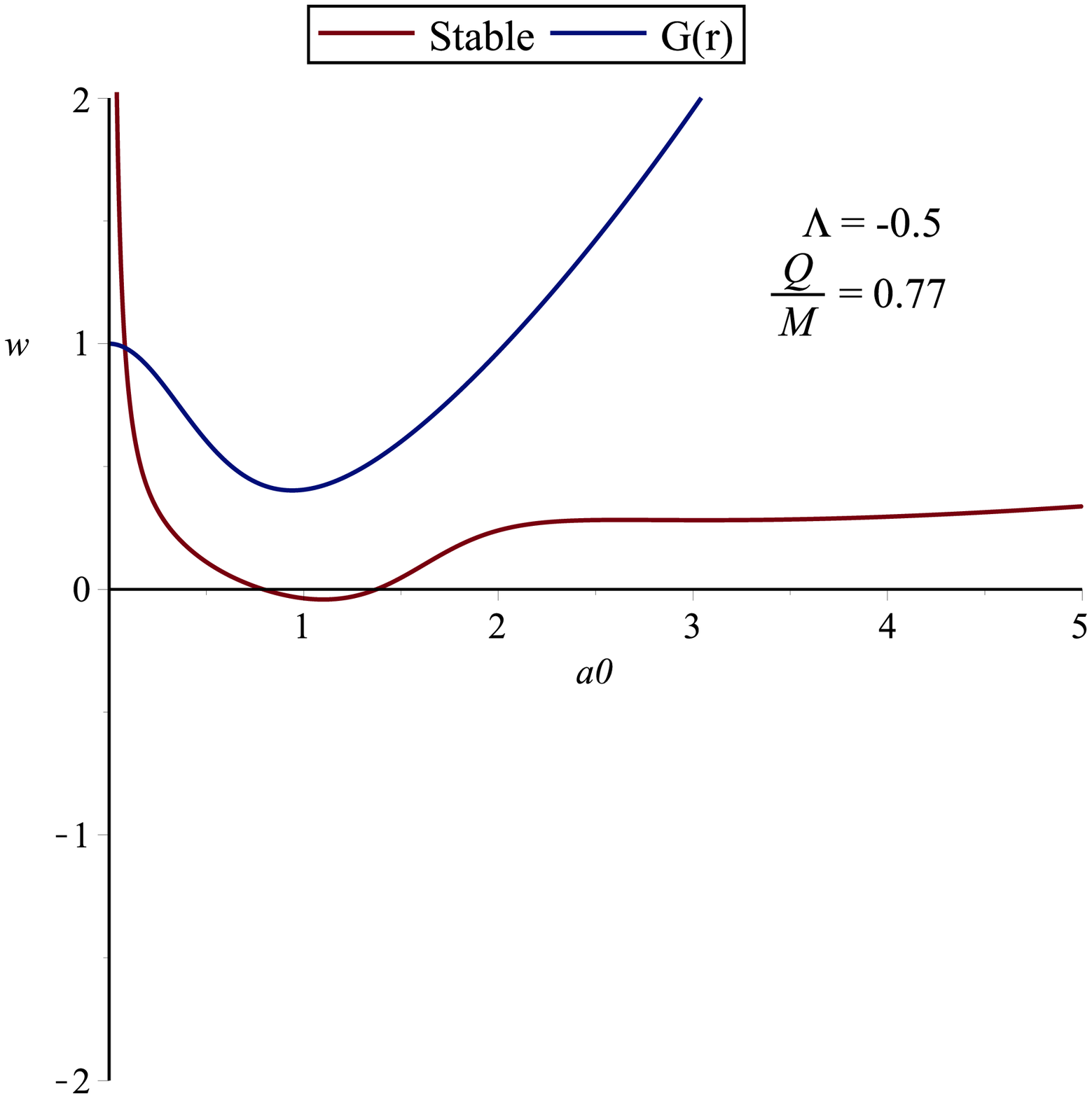,
width=0.45\linewidth}\caption{Plots for stable regular ABG wormholes
with logarithmic gas EoS and $\Lambda=0.1, -0.5$.}
\end{figure}

\section{Stability Analysis under Velocity Perturbations}

Now, we study stability of regular ABG thin-shell wormholes against
velocity perturbations. We take small velocity perturbations about
static configuration $a=a_{0}$ such that we can consider an
approximately static fluid for exotic matter after any perturbation.
This fact leads to an assumption that one can take dynamic EoS for
wormhole same as that of static EoS \cite{10}. In this context, we
use Eq.(\ref{12}) which leads to
\begin{equation}\label{34}
p=-\frac{1}{2}\left(1+\frac{aG'(a)}{2G(a)}\right)\sigma.
\end{equation}
Inserting $\sigma$ and $p$ from Eqs.(\ref{9}) and (\ref{10}), it
follows that
\begin{equation}\label{35}
\ddot{a}-\frac{G'(a)}{2G(a)}\dot{a}=0,
\end{equation}
which corresponds to the wormhole's throat motion in one-dimension.
Integration of this equation yields
\begin{equation}\label{36}
\dot{a}=\dot{a_{0}}\frac{\sqrt{G(a)}}{\sqrt{G(a_{0})}},
\end{equation}
whose second integration gives
\begin{equation}\label{37}
\frac{\dot{a_{0}}}{\sqrt{G(a_{0})}}(\tau-\tau_{0})=\int_{a_{0}}^a
\frac{da}{\sqrt{G(a)}}.
\end{equation}
It is mentioned here that $\dot{a}_{0}$ is assumed to be non-zero
initial small velocity of throat. In the following, we provide
examples to analyze stability under small velocity perturbations.

\subsection {The Schwarzschild Case}

For Schwarzschild metric function $G(a)=1-\frac{2M}{a}$,
Eq.(\ref{37}) yields
\begin{equation}\label{38}
\frac{\dot{a_{0}}}{\sqrt{G(a_{0})}}(\tau-\tau_{0})=a\sqrt{G(a)}-
a_{0}\sqrt{G(a_{0})}+M\ln\left(\frac{a-M+a\sqrt{G(a)}}
{a_{0}-M+a_{0}\sqrt{G(a_{0})}}\right),
\end{equation}
which shows the non-oscillatory motion. Thus, the wormhole throat
for Schwarzschild case will be unstable under small velocity
perturbations.

\subsection {The ABG Case}

For the regular ABG BH, Eq.(\ref{37}) upto the third order of $Q$
gives
\begin{eqnarray}\nonumber
\frac{\dot{a_{0}}}{\sqrt{G(a_{0})}}(\tau-\tau_{0})&=&a\sqrt{G(a)}-
a_{0}\sqrt{G(a_{0})}+M\ln\left(\frac{a-M+a\sqrt{G(a)}}
{a_{0}-M+a_{0}\sqrt{G(a_{0})}}\right)\\\label{39}&-&\frac{(a-a_{0})^3}{6}\left
(\frac{1}{Q^2}-\frac{2M}{Q^3}\right),
\end{eqnarray}
which again indicates the non-oscillating motion. Consequently, the
wormhole throat for regular ABG spacetime will remain unstable under
velocity perturbations. Also, Eq.(\ref{35}) provides the throat's
acceleration $\ddot{a}=\frac{G'(a)}{2G(a)}\dot{a}$ to be positive
leading to unstable ABG wormhole configurations.

\section{Conclusions}

This paper is devoted to construct regular ABG thin-shell wormholes
by implementing the Visser's cut and paste technique. We have
formulated the surface stresses by using Lanczos equation which
indicate the existence of exotic matter due to the violation of NEC
and WEC. It is found that the wormhole has attractive or repulsive
characteristics for $a^r>0$ and $a^r<0$, respectively. Moreover, we
have quantified the total amount of exotic matter by the volume
integral theorem which shows that a small quantity of exotic matter
is required to support the wormhole. We have analyzed stability of
the constructed thin-shell wormholes by incorporating the effects of
increasing values of electric charge. We have taken $a>r_{h}$ in
order to neglect the presence of event horizons for wormhole's
viability.

We have considered perturbation of the form $p=\Phi(\sigma)$ and
studied stability conditions graphically for $\Delta''>0$. To find
any realistic dark energy candidate, we have chosen linear,
logarithmic, CG, GCG and MGCG models as exotic matter at the
wormhole throat. The trivial case $\frac{Q}{M}=0$ corresponds to
Schwarzschild spacetime. We have investigated stable regions within
a physically acceptable range of different parameters (Figures
\textbf{4}-\textbf{13}). We have also plotted the function $G(r)$ to
evaluate the position of wormhole throat and event horizon of BH,
where we have choosen $a_{0}=r$ at thin-shell (hypersurface
$\Sigma$). It is observed that the stability areas for linear gas
tend to decrease by increasing the values of $\frac{Q}{M}$ while CG
shows the least stable regions for the respective wormhole
solutions. For GCG and MGCG, we have analyzed more stability regions
which decrease gradually with increasing $\frac{Q}{M}$ and diminish
to only one stable region for $\frac{Q}{M}=1.1$. It is found that
the role of logarithmic gas is to enhance stable regions for regular
ABG wormhole configurations. We have concluded that the wormhole
stability highly depends on the values of electric charge.

We have also investigated the stable regions for regular ABG
thin-shell wormholes in de Sitter $(\Lambda=0.1)$ and anti-de Sitter
$(\Lambda=-0.5)$ backgrounds. In this context, linear gas shows more
stable regions as compared to the regions for anti-de Sitter
spacetime. It is observed that GCG gives maximum stability areas for
regular ABG thin-shell wormholes corresponding to different values
of $\frac{Q}{M}$ with $\Lambda<0$. For MGCG and logarithmic gas, it
is found that the increasing value of $\frac{Q}{M}$ shows
enlargement in the stable regions and maximum stable regions are
obtained for $\Lambda=-0.5$. We have also discussed small velocity
dependent perturbations and analyzed that the constructed wormhole
configurations will be unstable under these perturbations. A
consequence of this instability is that the wormhole throat may
expand or collapse to a BH depending on the signs of velocity. It is
observed that there are more stable wormhole solutions in de Sitter
and anti-de Sitter spacetimes in comparison to the case without
$\Lambda$. We conclude that stability regions may expand/shrink by
tuning the values of charge and other parameters.

\end{document}